\documentclass[cernpreprint,atlasdraft=false,texlive=2017,coverpage=false,USenglish,texmf]{atlasdoc}
 
\usepackage{atlaspackage}
\usepackage{atlasbiblatex}
 
\usepackage{atlasphysics}
\usepackage{atlasprocess}
\usepackage{atlasjetetmiss}
\usepackage{atlasmisc}
\usepackage{multirow}
\usepackage{tikz-feynman}
\usetikzlibrary{positioning,automata}
 
\usetikzlibrary{positioning,arrows.meta}
\tikzfeynmanset{ with arrow/.style = {
decoration={
markings,
mark=at position 0.5
with {\arrow[xshift=2mm]{Stealth[black,width=1mm,length=2mm]}}
},
postaction=decorate}
}

\graphicspath{{logos/}{figures/}}
 
\usepackage{ANA-EXOT-2018-16-PAPER-defs}
 
\usepackage{PMGRefs-defs}
 
\usepackage{lscape}

\AtlasTitle{Search for New Phenomena in Final States with Two Leptons and One or No $b$-Tagged Jets at $\sqrt{s} = 13$ TeV Using the ATLAS Detector}

\AtlasJournalRef{Phys. Rev. Lett. 127 (2021) 141801}
\AtlasDOI{10.1103/PhysRevLett.127.141801}
\PreprintIdNumber{CERN-EP-2021-065}
 
\AtlasAbstract{
A search for new phenomena is presented in final states with two leptons and one or no $b$-tagged jets.
The event selection requires the two leptons to have opposite charge, the same flavor (electrons or muons), and a large invariant mass.
The analysis is based on the full run-2 proton-proton collision dataset recorded at a center-of-mass energy of $\sqrt{s} = 13$~TeV by the ATLAS experiment at the LHC, corresponding to an integrated luminosity of 139~fb$^{-1}$.
No significant deviation from the expected background is observed in the data.
Inspired by the $B$-meson decay anomalies, a four-fermion contact interaction between two quarks ($b,s$) and two leptons ($ee$ or $\mu\mu$) is used as a benchmark signal model, which is characterized by the energy scale and coupling, $\Lambda$ and $g_*$, respectively.
Contact interactions with $\Lambda / g_*$ lower than 2.0 (2.4)~TeV are excluded for electrons (muons) at the 95\% confidence level, still far below the value which is favored by the $B$-meson decay anomalies.
Model-independent limits are set as a function of the minimum dilepton invariant mass, which allow the results to be reinterpreted in various signal scenarios.
}
 
\author{The ATLAS Collaboration}

\hypersetup{pdftitle={ATLAS document},pdfauthor={The ATLAS Collaboration}}
 
\begin{document}
 
\maketitle

Lepton flavor universality (LFU) is one of the fundamental predictions of the standard model (SM).
LFU was tested extensively at LEP and SLD~\cite{ALEPH:2005ab} and found to be compatible with the SM prediction.
Recent measurements hint at a possible violation of LFU in rare $B$-meson decays~\cite{Aaij:2017vbb,Aaij:2019wad,Aaij:2019bzx,Lees:2012tva,Abdesselam:2019wac,Abdesselam:2019lab,Aaij:2017vbb,Aaij:2014pli,Aaij:2015esa,Aaij:2015xza,Aaij:2016flj,Aaij:2020nrf,Aaij:2015oid} into a $K$ meson and a pair of muons or electrons.
Possible extensions to the SM suggest that the decay mechanism implies that physics beyond the SM (BSM) is present between the initial ($b$ quark) and final states ($s$ quark and two charged electrons or muons).
The BSM interaction can be modeled using an effective field theory (EFT) with a four-point contact interaction between the fermions involved ($b s \ell \ell$, $\ell = e, \mu$),
where the scale and coupling of the underlying physics are denoted by $\Lambda$ and $g_*$, respectively.\footnote{For a discussion regarding the validity of the EFT parameter space see e.g.\ Ref.~\cite{Afik:2019htr}.} It can be searched for in final states with two opposite-charge and same-flavor leptons produced in association with exactly one $b$ quark or without any $b$ quarks.
To explain the asymmetries measured in the $B$-meson decays, the $b s \ell \ell$ interaction would have to be different between electrons and muons.
The phenomenological framework for this analysis was suggested in Ref.~\cite{Afik:2018nlr}.
The $B$-meson decay anomalies could correspond to a $b s \ell \ell$ operator with $\Lambda / g_* \approx 30$~\TeV~\cite{Alguero:2019ptt,Greljo:2017vvb}, which is beyond the discovery reach of the present search.
However, this unique signature may provide enhanced sensitivity to other signal scenarios as well~\cite{Afik:2019htr,Afik:2021jjh}.
Fig.~\ref{fig:Feynman_bsll_SM} shows Feynman diagrams for $B$-meson decays, via the SM and via a $b s \ell \ell$ contact interaction, and for the production process via a $b s \ell \ell$ contact interaction in proton-proton ($pp$) collisions.\footnote{At tree level, there is no SM amplitude, and therefore interference effects are neglected.}
 
In this Letter, a search for new phenomena is presented, using $pp$ collisions at the Large Hadron Collider (LHC) with a center-of-mass energy of \sqsthirt .
Data recorded by the ATLAS detector~\cite{PERF-2007-01} during 2015--2018 are used, corresponding to an integrated luminosity of 139~\ifb.
Final states with two oppositely charged electrons or muons are considered separately, and further categorized into events with either no $b$-tagged jets or exactly one $b$-tagged jet.
The \bsll{} EFT~\cite{Afik:2018nlr} is considered as a benchmark model, and model-independent results are also presented.

ATLAS is a multipurpose particle detector with a forward-backward symmetric cylindrical geometry and a near $4\pi$ coverage in a solid angle.\footnote{ATLAS uses a right-handed coordinate system with its origin at the nominal interaction point (IP) in the center of the detector and the $z$-axis along the beam pipe. The $x$-axis points from the IP to the center of the LHC ring, and the $y$-axis points upwards. Cylindrical coordinates ($r$, $\phi$) are used in the
transverse plane, $\phi$ being the azimuthal angle around the $z$-axis. The pseudorapidity is defined in terms of the polar angle $\theta$ as $\eta =  -\ln\tan(\theta/2)$. Angular distance is measured in units of $\Delta R \equiv \sqrt{(\Delta\eta)^2 + (\Delta\phi)^2}$. The transverse momentum is defined as $\pT = p \sin\theta = p/\cosh \eta$, and the transverse energy, \ET , is defined analogously.}
It consists of an inner tracking detector surrounded by a thin superconducting solenoid providing a \SI{2}{\tesla} axial magnetic field, electromagnetic and hadronic calorimeters, and a muon spectrometer.
The inner tracking detector (ID) covers the pseudorapidity range $|\eta| < 2.5$.
It consists of silicon pixel, silicon microstrip, and transition radiation tracking detectors, and a new innermost $B$ layer, added to the pixel detector before run~2~\cite{ATLAS-TDR-2010-19,PIX-2018-001}.
Lead and liquid-argon (LAr) sampling calorimeters provide electromagnetic (EM) energy measurements with high granularity.
A steel and scintillator-tile hadronic calorimeter covers the central pseudorapidity range ($|\eta| < 1.7$).
The end cap and forward regions are instrumented with LAr calorimeters for EM and hadronic energy measurements up to $|\eta| = 4.9$.
The muon spectrometer (MS) surrounds the calorimeters and is based on three large air-core toroidal superconducting magnets with eight coils each.
The field integral of the toroids ranges between \num{2.0} and \SI{6.0}{\tesla\metre} across most of the detector.
The muon spectrometer includes a system of precision tracking chambers and fast detectors for triggering.

\begin{figure}
\centering
\resizebox{\textwidth}{!}{
\includegraphics[width=0.90\textwidth]{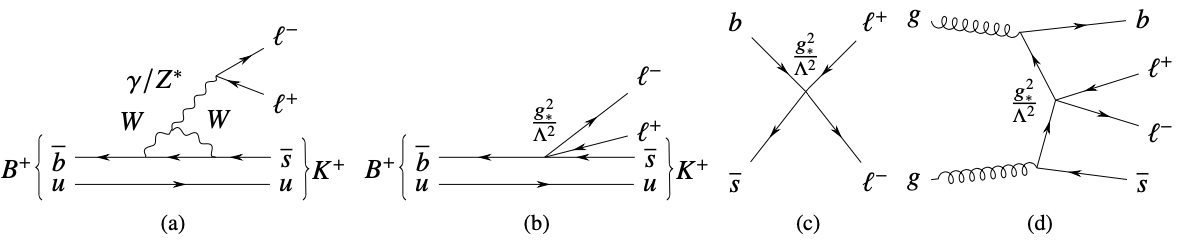}
}
\caption{Representative Feynman diagrams for the decay of a $B^{+}$ meson to a $K^{+}$ meson in association with two leptons (a) in the SM and (b) in the EFT approach, and for production of two leptons via a $b s \ell \ell$ contact interaction in $pp$ collisions (c) without and (d) with a $b$ jet in the final state.}
\label{fig:Feynman_bsll_SM}
\end{figure}

Monte Carlo (MC) simulations are used to model the expected SM background and the benchmark signals.
All background and signal MC samples were generated using the five-flavor scheme.
The \POWHEGBOX~[v1] MC generator~\cite{Nason:2004rx,Frixione:2007vw,Alioli:2010xd,Alioli:2008gx} was used to simulate at next-to-leading order (NLO) in QCD the inclusive hard-scattering $Z / \gamma^* \to \ell^+ \ell^-$ sample, denoted as $Z / \gamma^*$+jets, using the CT10 parton distribution function (PDF) set~\cite{Lai:2010vv}.
It was interfaced to \PYTHIA~[8.186] to model the parton shower, hadronization, and underlying event, using the AZNLO tune~\cite{STDM-2012-23} and the CTEQ6L1 PDF set~\cite{Pumplin:2002vw}.
The $Z / \gamma^*$+jets samples were normalized to next-to-next-to-leading order (NNLO) in QCD and corrected for remaining NLO electroweak effects following the procedure described in Ref.~\cite{EXOT-2016-05}.
The effect of QED final-state radiation (FSR) was simulated with \PHOTOS++~3.52~\cite{Golonka:2005pn,Davidson:2010ew}.
The use of \POWHEGBOX was validated by a generator-level comparison with a sample produced by
\SHERPA~[2.2.1]~\cite{Bothmann:2019yzt}
using NLO matrix elements for up to two partons and leading-order (LO) matrix elements
for up to four partons calculated with the Comix~\cite{Gleisberg:2008fv}
and \openloops~\cite{Buccioni:2019sur,Cascioli:2011va,Denner:2016kdg} libraries.
Samples of diboson ($W$-boson) events, denoted by $VV$ ($W+$jets), were simulated with \SHERPA~[2.2.2~(2.2.1)]~\cite{Bothmann:2019yzt} using the \nnpdfnnlo PDF set, with matrix elements at NLO in QCD with up to one (two) additional partons and up to three (four) additional parton emissions at LO~\cite{Gleisberg:2008fv,Buccioni:2019sur,Cascioli:2011va,Denner:2016kdg}. 
For both $VV$ and $W+$jets, the matrix elements were matched with the \SHERPA parton shower~\cite{Schumann:2007mg} using the MEPS@NLO prescription~\cite{Hoeche:2011fd,Hoeche:2012yf,Catani:2001cc,Hoeche:2009rj} and the parameter tune developed by the \SHERPA authors.
The $W+$jets samples were normalized to a NNLO prediction~\cite{Anastasiou:2003ds}.
The production of \ttbar\ and single-top-quark $Wt$ events was modeled using the \POWHEGBOX~[v2] generator at NLO with the \nnpdfnlo PDF set and the \hdamp\ parameter set to $1.5 m_\mathrm{top}$.
Events were passed to \PYTHIA~[8.230]~\cite{Sjostrand:2014zea} to model the parton shower, hadronization, and underlying event, using the A14 parameter tune~\cite{ATL-PHYS-PUB-2014-021} and the \nnpdftwo PDF set.
For $Wt$ events, the diagram removal scheme~\cite{Frixione:2008yi} was used to eliminate interference with \ttbar\ production.
The production of \ttV\ events was modeled using the \mgamc~v2.3.3~\cite{Alwall:2014hca} generator at NLO with the \nnpdfnlo PDF set.
The events were interfaced to \PYTHIA~[8.210] using the A14 tune and the \nnpdftwo PDF set.
The \evtgen~1.2.0 (1.6.0) program~\cite{Lange:2001uf} was used to decay bottom and charm hadrons for the \ttV{} and $Z / \gamma^*$+jets (\ttbar) processes.
The $b s \ell \ell$ EFT signal was generated at LO, using a model provided by the authors of Ref.~\cite{Afik:2018nlr},\footnote{The model can be found at the FeynRules database: \href{http://feynrules.irmp.ucl.ac.be/wiki/FCNC4F}{\textcolor{blue}{http://feynrules.irmp.ucl.ac.be/wiki/FCNC4F}}.} with up to two partons in the final state by \mgamc with the \nnpdftwo PDF set and the A14 tune of \PYTHIA~[8] parameters.
The CKKW-L merging algorithm~\cite{Lonnblad:2011xx} was used with a \kt-Durham parameter of 400~\GeV{}.
The cross section for the simulated signal with $\Lambda/g_* = 1$~\TeV{} is 0.113~pb, for both electrons and muons.
The ATLAS detector response was simulated with \textsc{Geant4}~\cite{Agostinelli:2002hh, SOFT-2010-01}, except for signal samples, where a fast simulation~\cite{ATL-PHYS-PUB-2010-013} was used for the calorimeter response and \textsc{Geant4} for all other detector systems.
The effect of multiple interactions in the same and neighboring bunch crossings (pileup) was modeled by overlaying simulated inelastic $pp$ events generated by \PYTHIA~[8.186]~\cite{Sjostrand:2007gs} with the A3 tune~\cite{ATL-PHYS-PUB-2016-017} and the \nnpdftwo PDF set~\cite{Ball:2014uwa}.
The MC distributions were reweighted to the distribution of the average number of interactions per bunch crossing in data.

Only events taken during stable beam conditions, and for which all relevant components of the detector were operational, are considered.
Single-lepton triggers were used~\cite{TRIG-2018-05,TRIG-2018-01}, with \pT{} threshold of 60~\GeV{} or 140~\GeV{} for electrons, depending on the identification requirement, and 50~\GeV{} for muons.
Events must have a vertex with at least two tracks with a minimum \pT of 500~\MeV, where the highest $\Sigma_{\mathrm{tracks}} p_{\mathrm{T}}^{2}$ vertex is chosen as the primary one~\cite{ATL-PHYS-PUB-2015-026}.
 
Electrons are reconstructed from energy clusters in the EM calorimeter with ID tracks matched to them and are required to fulfill the ``tight likelihood'' identification criteria as well as calorimeter- and track-based isolation criteria~\cite{EGAM-2018-01}.
Electrons must have a minimum transverse energy of 30~\GeV{} and must be within the region $|\eta_{\mathrm{cluster}}|$~<~2.47, excluding the transition region between the barrel and the end cap, 1.37~<~$|\eta_{\mathrm{cluster}}|$~<~1.52.
Muons are reconstructed from combined MS and ID tracks with a minimum \pT{} of 30~\GeV{}, must fulfill the ``high-\pT{}'' identification criteria~\cite{MUON-2018-03}, which aim to optimize the momentum resolution for tracks with high transverse momentum, and must be within the region $|\eta|$~<~2.5.
For muons, track-based isolation criteria are required based on the scalar sum of the transverse momenta of the ID tracks associated with the primary vertex, excluding the muon track itself.
Muon (electron) candidates are required to originate from the primary vertex by requiring the significance of the track's transverse impact parameter calculated relative to the beam line $d_0 / \sigma(d_0)$ to be smaller than 3.0 (5.0). Furthermore, the longitudinal impact parameter $\zzero$, defined as the difference between the $z$ coordinate of the point of closest approach to the beam line and the longitudinal position of the primary vertex, is required to satisfy $|\zzsth| < 0.5$~mm.
Anti-$k_t$ jets~\cite{Cacciari:2008gp} are reconstructed from energy deposits in topological clusters of calorimeter cells~\cite{PERF-2014-07}, using the particle-flow algorithm~\cite{PERF-2015-09} and a radius parameter of 0.4.
The jet energy is calibrated at particle level~\cite{JETM-2018-05}.
Jets are required to be within $|\eta|$~<~2.5 and to have a minimum \pT of 30~\GeV.
A jet vertex tagger~\cite{PERF-2014-03} is used to suppress pileup contributions for jets with $|\eta|$~<~2.4 and \pT~<~60~\GeV.
Jets are identified as containing $b$ hadrons using the DL1 algorithm~\cite{ATL-PHYS-PUB-2017-013,PERF-2012-04,ATL-PHYS-PUB-2017-013}, with a $b$-tagging efficiency of $\sim$77\% for $b$ jets, and a rejection factor of $\sim$6 for $c$ jets and $\sim$110 for other light jets, based on simulated \ttbar events.
Finally, a sequential overlap-removal procedure is used as follows:
in the first step, electrons that share a track with a muon are removed from the event;
in the second step, any jet that has a $\DeltaR$ to an electron that is smaller than 0.2 is removed from the event; and in the third step, electrons are removed from an event if they are geometrically closer than $\DeltaR = 0.4$ to any remaining jet.
Jets within $\DeltaR < 0.04 + 10~\GeV{}/\pT(\mu)$ to a muon are removed from the event if they have, at most, two associated tracks with $\pT(\mathrm{track}) > 0.5~\GeV$, otherwise the muon is removed.

Events are selected by requiring two same-flavor electrons or muons with opposite electric charge, where at least one of the leptons is required to geometrically match the object that fired the trigger.
To ensure high trigger efficiency, the \pT threshold for the leading lepton is raised to 65~\GeV{}.
Two categories are defined depending on the presence of a $b$-tagged jet, targeting two different production mechanisms.
The $b$-veto category, denoted by $e^+ e^- / \mu^+ \mu^- + 0b$, discards any event with a $b$-tagged jet, while the $b$-tag category, denoted by $e^+ e^- / \mu^+ \mu^- + 1b$, requires exactly one $b$-tagged jet in each event.
No further requirement on the number of jets is made.
Regions in each category are defined based on the dilepton invariant mass \mll{} and are selected to allow high statistics to constrain the dominant backgrounds in dedicated control regions (CRs), validate the background estimation in dedicated validation regions (VRs), and keep a broad set of signal regions (SRs).
SRs are defined with lower bounds on \mll{}, $m_{\ell \ell}^\mathrm{min}$, ranging from 400~\GeV{} to 3200 (2000)~\GeV{} for the $b$-veto ($b$-tag) category with a step size of 100~\GeV{}, where each SR is defined by requiring $m_{\ell \ell} > m_{\ell \ell}^\mathrm{min}$.
CRs are defined in order to normalize the contribution of the two dominant background processes originating from \ttbar{}, $Wt$ and \ttV{}, together denoted by ``top,'' and $Z / \gamma^*$+jets processes.
The $Z / \gamma^*$+jets CRs ($Z$-CRs) are defined by requiring events to be within 130~\GeV{}~<~\mll{}~<~250~\GeV{},
while the intermediate mass range, 250~\GeV{}~<~\mll{}~<~400~\GeV{}, serves as a VR to test the background modeling.
For each $Z$-CR and VR the same $b$-veto and $b$-tag categories as in the SRs are applied.
Finally, a top-CR is constructed by requiring exactly two $b$-tagged jets and the dilepton invariant mass to satisfy \mll{}~>~130~\GeV{}.

A fit-based extrapolation procedure is used to estimate the tails of the top \mll{} distributions, which suffer from low statistics in the MC simulation, using functions developed in other ATLAS searches~\cite{EXOT-2018-30},
\begin{equation}\nonumber
f^{\mathrm{bkg1}}(\mll) = \mathrm{e}^{-a} \mll^{b} \mll^{c \log{(\mll)}}\;~\mathrm{and}~\; f^{\mathrm{bkg2}}(\mll)  = \frac{a}{\left(\mll+b\right)^{c}},
\end{equation}
where $a,b$ and $c$ are free parameters.
Several fits are performed by using both functions, while varying the start and end point of the fit range and using a $\chi^2$ test to estimate the level of agreement between the fits and the MC prediction.
The fit with the lowest $\chi^2$ provides the nominal choice of the function parameter values, while all other fits with $\chi^2$ probability smaller than a fixed $\chi^2$ value are used for the uncertainty estimation.
This fixed $\chi^2$ value is chosen such that, near the transition point between the simulation and the extrapolation, the resulting uncertainty on the extrapolation is similar to the overall uncertainty, which is accounting for the experimental and modeling systematic uncertainty, and the statistical uncertainty of the simulated top background samples.
Furthermore, checks are performed in order to make sure that the fitted function reproduces the MC event yields at lower values of $m_{\ell \ell}$ and that the cumulative distribution of the extrapolation is consistent with the integrated event yields in the MC samples.
Finally, since the extrapolation is done for the combined top sample, which includes all top-related processes, it was checked that those processes have a similar $m_{\ell \ell}$ shape within uncertainties.
For the top background extrapolation, the transition points between simulation and extrapolation in the \mll{} distributions are (1000, 1200, 1200 or 1300)~\GeV{} for the $(0, 1, 2)$-$b$-tagged jets selections, respectively, in the electron or muon channel.
Above the transition point, only the extrapolation uncertainty is assigned to the top background sample.
This uncertainty is the dominant one in the $b$-tag categories. It is $46\%$ ($53\%$) and $223\%$ ($236\%$) relative to the nominal fitted extrapolation in the $e^+ e^- + 1b$ ($\mu^+ \mu^- + 1b$) category with $m_{\ell \ell}^\mathrm{min} =$1200 and 2000~\GeV{}, respectively.

The background contribution of events with reconstructed objects that have been misidentified as leptons, referred to as ``multijet,'' is estimated using a data-driven approach in the electron channel.
In the muon channel, this contribution is found to be negligible.
The matrix method is used, similar to the procedure described in Ref.~\cite{EXOT-2016-05}.
The probabilities that a jet and a real electron satisfy the electron identification criteria are evaluated, for both the nominal and the ``loose likelihood'' identification criteria, while for the former no isolation criteria are applied.
Then, these probabilities are used in order to estimate the multijet contribution in the selected region.
The multijet background estimation suffers from low statistics at high $m_{\ell \ell}$, and an extrapolation procedure similar to that of the top processes is used, with
transition points at (800, 600, 600)~\GeV{} for the $(0,1,2)$-$b$-tagged jets selections, respectively.

Experimental systematic uncertainties, related to the modeling of the detector response in the simulation, are considered.
The uncertainty in the combined 2015--2018 integrated luminosity is 1.7\%~\cite{ATLAS-CONF-2019-021}.
Uncertainties in electron and muon trigger, reconstruction, and identification efficiencies, and energy and momentum calibration and resolution, are derived from data using \Zll{} and $J / \psi \to \ell \ell$ decays~\cite{EGAM-2018-01,PERF-2015-10}.
Uncertainties in the jet energy scale and resolution are evaluated from MC simulations and from data using multijet, \Zjets{}, and \Gjets{} events~\cite{JETM-2018-05}.
Uncertainties in the $b$-tagging efficiency are derived from data~\cite{FTAG-2018-01} for $b$ jets, $c$ jets, and other light jets.
MC simulations are used to extrapolate the efficiencies to regions beyond the kinematic reach of each calibration.
In order to assess the systematic uncertainty due to pileup, the reweighting to match simulation to data is varied within its uncertainty.
Finally, uncertainties related to the top and multijet background extrapolation are evaluated as described earlier in the text.

Theoretical systematic uncertainties, related to the modeling of the background processes in the MC simulation, are considered as well.
The $Z / \gamma^*$+jets PDF variation uncertainty is estimated using the 90\% confidence level (C.L.) CT14nnlo PDF error set, following Refs.~\cite{EXOT-2012-23,EXOT-2016-05,Gao:2013bia,Butterworth:2015oua}.
The uncertainty due to $\alpha_\mathrm{s}$ is assessed by using the CT14nnlo PDF set where the value of $\alpha_\mathrm{s}(m_Z) = 0.118$ is shifted by 0.003, while QCD scale uncertainties are obtained by varying the renormalization and factorization scales simultaneously by a factor of 2 up and down.
The uncertainty due to the choice of PDF set is estimated by using the NNPDF3.0 PDF set instead of the nominal choice of CT14nnlo~\cite{Butterworth:2015oua}.
Corrections due to photon-induced processes are estimated using the MRST2004qed PDF set~\cite{Martin:2004dh}.
The uncertainty due to NLO electroweak corrections for the $Z / \gamma^*$+jets sample are evaluated as in Ref.\cite{EXOT-2012-23}.
For \ttbar{} and single-top-quark production, an uncertainty in the cross section originating from scale, PDF+$\alpha_\mathrm{s}$ and top-quark-mass uncertainties is applied.
The nominal sample is compared with a sample generated with \MGNLO to estimate the matrix-element uncertainty.
To evaluate the parton-shower uncertainty, a sample simulated with \POWHEGBOX interfaced to \HERWIG~[7]~\cite{Bellm:2015jjp} is used.
To simulate higher parton radiation, the factorization and renormalization scales are varied by a factor of 0.5 in the matrix element using the ``up'' variation from the A14 parameter tune in the parton shower.
For lower parton radiation, the renormalization and factorization scales are varied by a factor of 2.0 using the ``down'' variation in the parton shower.
The impact of FSR is evaluated by changing the renormalization scale for QCD emission by factors of 0.5 or 2.0.
For \ttbar{} and single-top-quark events, the PDF uncertainty is derived using 30 eigenvector variations as specified in Ref.~\cite{Butterworth:2015oua}, to estimate distribution shape uncertainties.
For \ttbar{} production, the impact of factorization and renormalization scale uncertainties on the shapes of distributions is derived by varying those scales by a factor of 0.5 or 2.0.
The nominal $Wt$ sample is compared with a sample generated using the diagram subtraction scheme~\cite{Frixione:2008yi,ATL-PHYS-PUB-2016-020}.
Finally, the statistical uncertainties of the simulated event samples are also taken into account.
 
Table~\ref{tab:sys:bsll} presents the systematic uncertainties for one signal region from each channel.
Systematic uncertainties that are lower than 0.5\% in a given region are not considered.

\begin{table}
\begin{center}
\centering
\caption{Summary of the relative systematic uncertainties for signal regions with $m_{\ell \ell}^\mathrm{min}=2000~(1500)$~\GeV{} before the fit is performed for the $0b$ ($1b$) categories. The background uncertainties are presented relative to the total SM prediction.}
\resizebox{\textwidth}{!}{
\begin{tabular}{|l|cc|cc|}
\hline
Source & \multicolumn{2}{c|}{$e^+ e^- + 0b~(1b)$ [\%]} & \multicolumn{2}{c|}{$\mu^+ \mu^- + 0b~(1b)$ [\%]} \\
& Signal $0b~(1b)$ & Background $0b~(1b)$ & Signal $0b~(1b)$ & Background $0b~(1b)$ \\
\hline
Luminosity & 1.7 (1.7) & 1.6 (1.5) & 1.7 (1.7) & 1.7 (1.7) \\
Pileup & $<$0.5 ($<$0.5) & $<$0.5 (0.7)~~\, & $<$0.5 ($<$0.5) & $<$0.5 ($<$0.5) \\
\hline
Leptons & 8.7 (8.6) & 8.6 (6.3) & 8.5 (6.5) & 9.1 (4.2) \\
Jets & $<$0.5 (1.8)~~\, & $<$0.5 (3.4)~~\, & $<$0.5 (1.6)~~\, & $<$0.5 (1.9)~~\, \\
$b$ tagging & $<$0.5 (1.4)~~\, & $<$0.5 (2.0)~~\, & $<$0.5 (1.4)~~\, & $<$0.5 (2.2)~~\, \\
\hline
Top bkg.\ extrapolation & - & ~~3.5 (32.0) & - & $<$0.5 (36.0)\, \\
Multijet extrapolation & - & ~~7.5 (15.0) & - & - \\
\hline
Top bkg.\ modeling & - & $<$0.5 ($<$0.5) & - & $<$0.5 ($<$0.5) \\
$Z / \gamma^*$+jets bkg.\ modeling & - & 9.4 (4.3) & - & 10.0 (5.5)~~ \\
\hline
MC statistics & 0.6 (0.8) & 1.9 (3.5) & 0.7 (1.0) & 1.7 (2.4) \\
\hline
Total & 8.9 (9.1) & 15.0 (37.0) & 8.7 (7.1) & 14.0 (37.0) \\
\hline
\end{tabular}
}
\label{tab:sys:bsll}
\end{center}
\end{table}

The signal and background yields are estimated using simultaneous maximum-likelihood fits of the signal-plus-background and background-only hypothesis.
Systematic and MC statistical uncertainties are included as nuisance parameters (NPs) and are constrained in the fit.
Dedicated fit parameters are used as additional NPs to adjust the top and $Z / \gamma^*$+jets background normalizations.
A likelihood ratio test statistic is used to assess the compatibility of the data with the background-only hypothesis to derive limits on the BSM signals, following the procedure in Ref.~\cite{Cowan:2010js}.
Exclusion limits are set using the CL$_\mathrm{s}$ method~\cite{Read:2002hq}, which is performed separately for each of the $b$-tag and $b$-veto categories in the electron and muon channels and by considering a single-bin SR and the relevant CRs per category.

The data agree well with the SM prediction in all of the VRs after the fit.
The postfit $m_{\ell \ell}$ distributions in the SRs are presented in Fig.~\ref{fig:limit:fitPlots:bsll:SR:mll:all} for the background-only hypothesis, while the fit is done only at the CRs (CR-only fit) and then used to estimate the background yields.
The cumulative $m_{\ell \ell}$ distribution for the signal regions after the CR-only fit to the data are shown in Fig.~\ref{fig:limit:fitPlots:bsll:SR:Cum:all} together with the yields in the different CRs and VRs.
The largest deviation from the SM prediction is observed in the $e^+ e^- + 1b$ category, where a selection of $m_{ee}^\mathrm{min} = 1700$~\GeV{} yields a local significance of $2.6\sigma$.
The global significance is estimated by generating pseudo-experiments using all of the electron $b$-tag SRs, and found to be $1.5\sigma$.
Other notable local deviations are in the $e^+ e^- + 1b$ category with $m_{ee}^\mathrm{min} = 1500,1600,2000~(1900)$~\GeV{}, which yields $2.1\sigma~(2.0\sigma)$, and in the $e^+ e^- + 0b$ category with $m_{ee}^\mathrm{min} = 2200$~\GeV{}, which yields $2.1\sigma$.
In the $\mu^+ \mu^- + 0b$ category, a deficit of events is observed with up to $1.9\sigma$, with a selection of $m_{\mu \mu}^\mathrm{min} = 1600,2800$~\GeV{}.
In Fig.~\ref{fig:limit:fitPlots:bsll:SR:limits:all}, model-independent upper limits on the signal cross section times selection efficiency times detector acceptance ($\sigma_{\textrm{vis}} = \sigma \cdot \epsilon \cdot \mathcal{A}$) are presented for each signal region selection.
For the \bsll{} benchmark model, the strongest expected limits are found with a selection of $m_{\ell \ell}^\mathrm{min}=1900~(1500)$~\GeV{} in the $e^+ e^- + 0b~(1b)$ category, which corresponds to expected and observed lower limits on $\Lambda / g_*$ of up to 2.2~(2.2)~\TeV{} and 2.0~(1.8)~\TeV{}, respectively, and with a selection of $m_{\ell \ell}^\mathrm{min}=1800~(1600)$~\GeV{} in the $\mu^+ \mu^- + 0b~(1b)$ category, which corresponds to expected and observed lower limits on $\Lambda / g_*$ of up to 2.1~(2.1)~\TeV{} and 2.4~(2.0)~\TeV{}, respectively.
The excluded values of $\Lambda / g_*$ are far below the value favored by the anomalies, which is $\approx 30$~\TeV.

\begin{figure}[!h]
\centering
\subfloat[]{\includegraphics[width=0.49\textwidth]{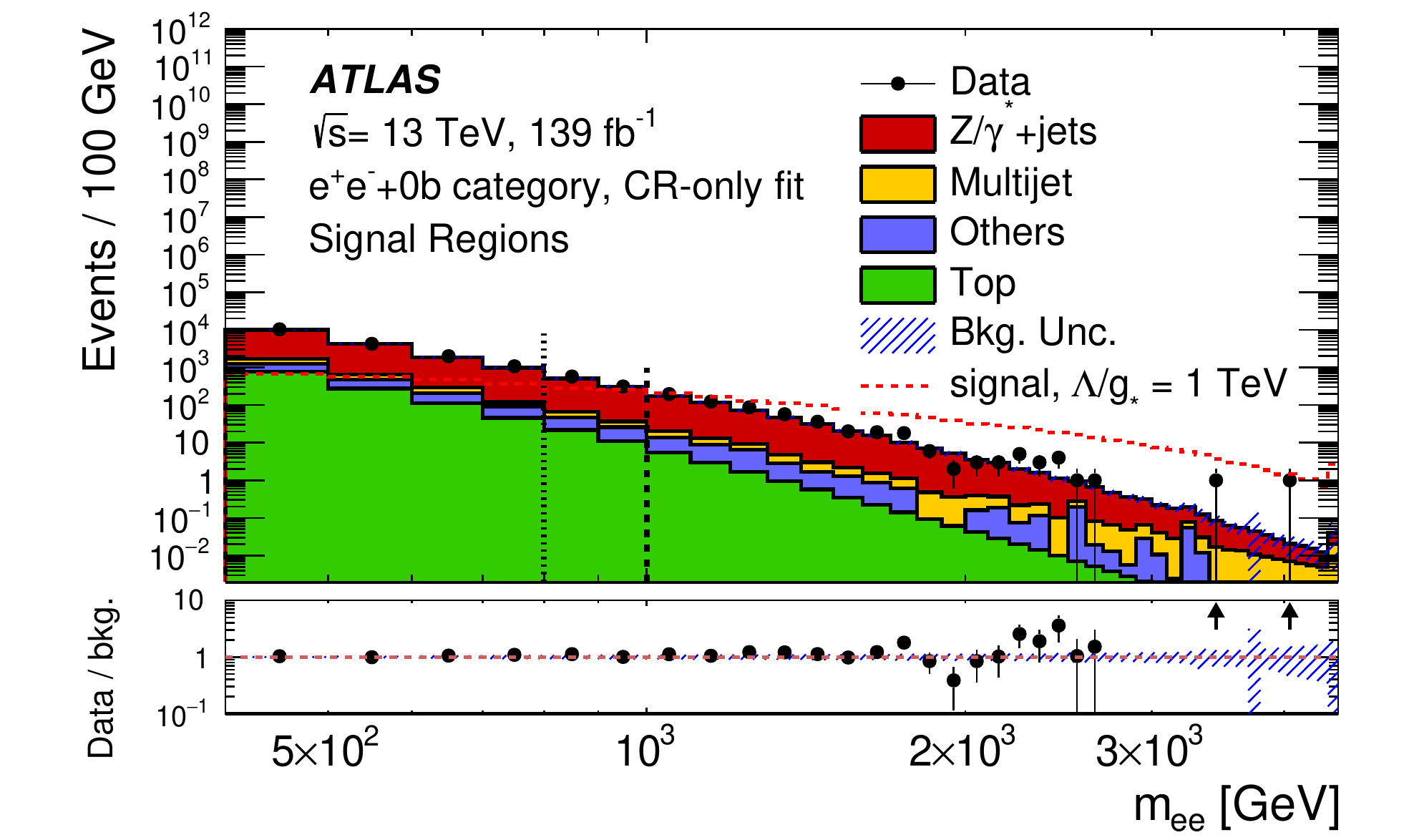}}
\subfloat[]{\includegraphics[width=0.49\textwidth]{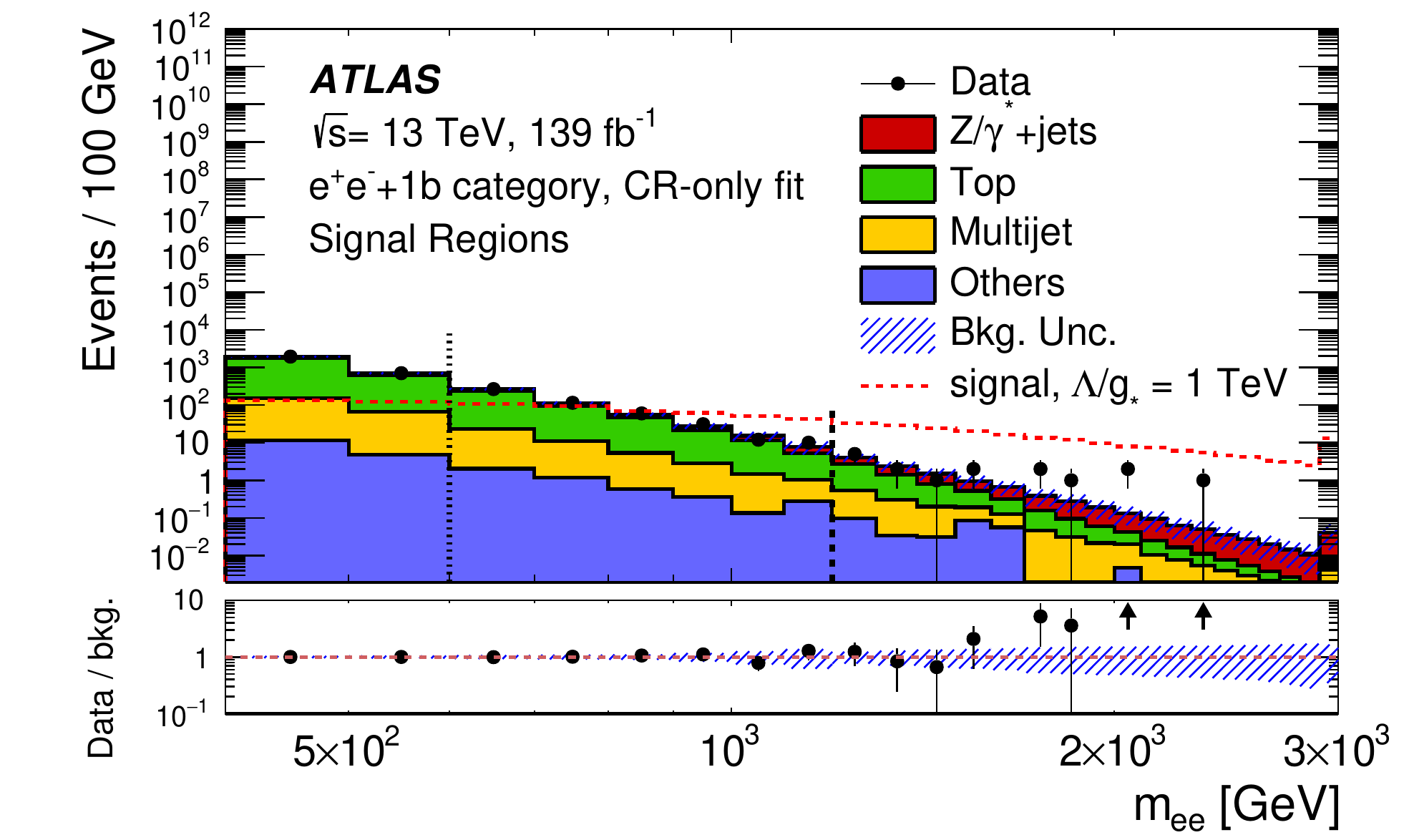}}\\
\subfloat[]{\includegraphics[width=0.49\textwidth]{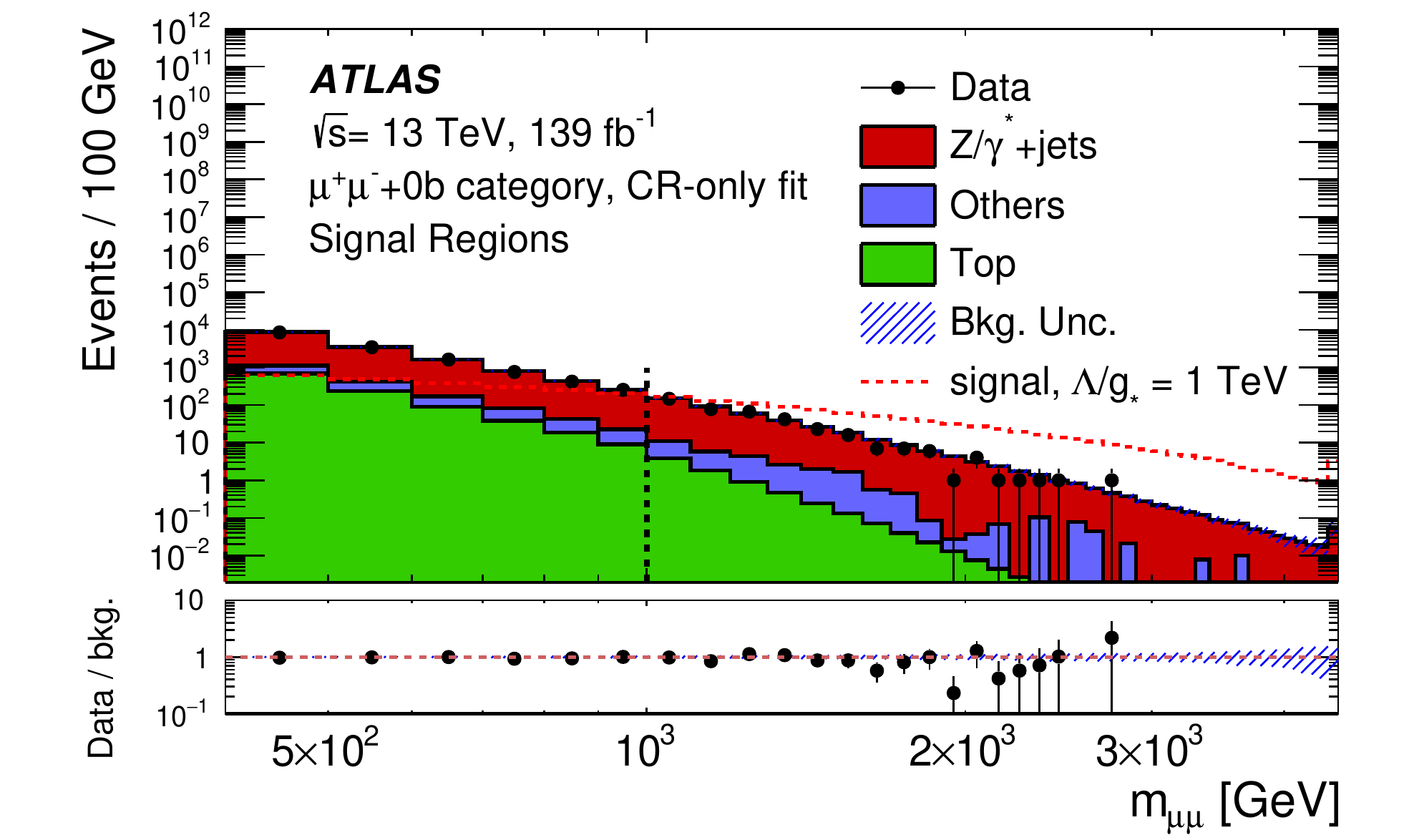}}
\subfloat[]{\includegraphics[width=0.49\textwidth]{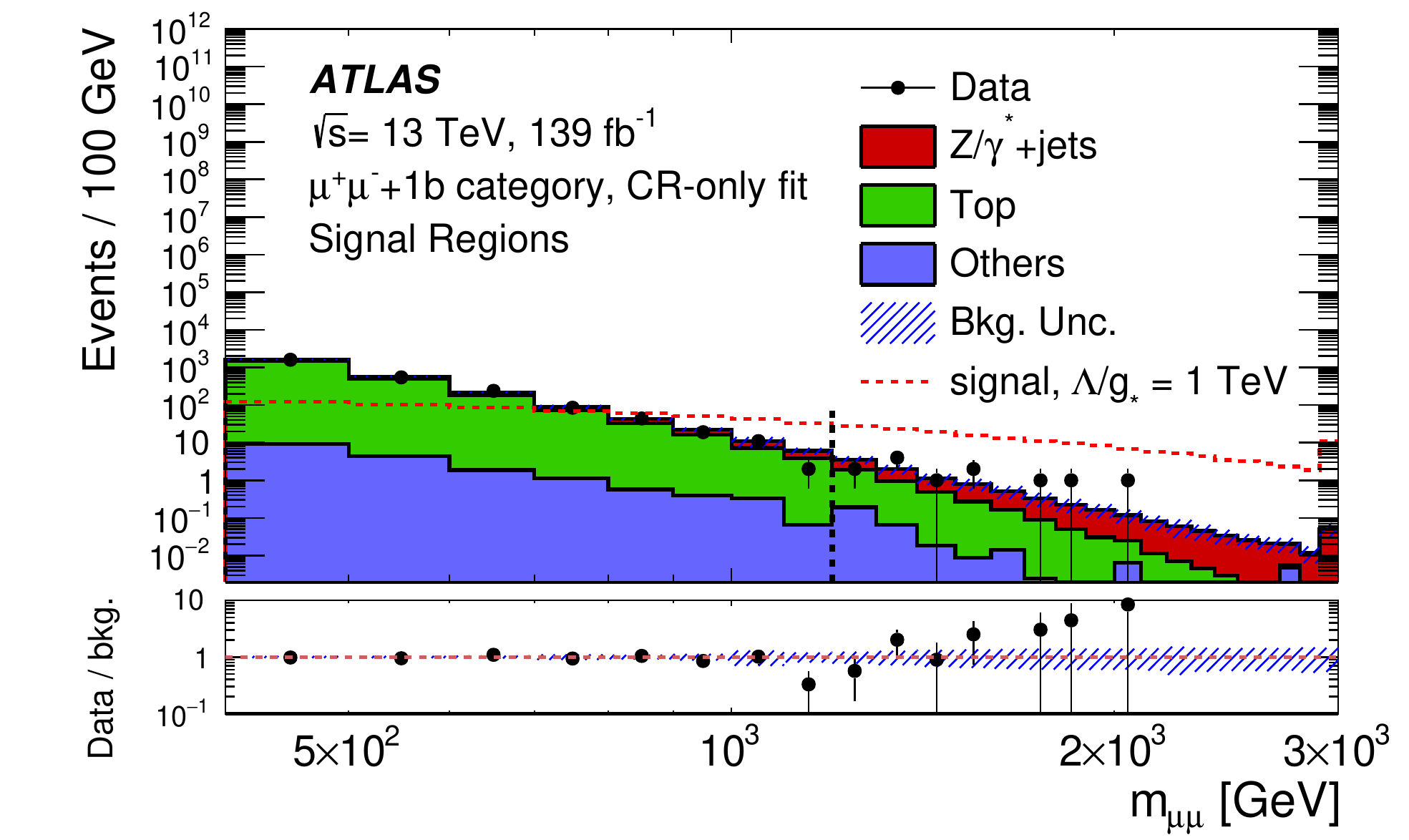}}\\
\caption{Data overlaid on SM background postfit $m_{\ell \ell}$ distributions in the SRs of the (a) electron $b$-veto, (b) electron $b$-tag, (c) muon $b$-veto and (d) muon $b$-tag categories. ``Others'' refers to diboson and $W+$jets events.
MC statistical uncertainties and systematic uncertainties are considered (hatched band).
The prefit signal distribution is presented as well for a hypothesis of $\Lambda / g_* = 1$~\TeV{}.
The bottom panels show the ratio of the data to the background prediction, while the arrows correspond to bins where the ratio is beyond the limits of the figure.
The last bin is an overflow bin, which contains the yields in the bins beyond it.
The dashed and dotted lines mark the transition point where the extrapolation is used in the analysis for the top and multijet backgrounds, respectively.
}
\label{fig:limit:fitPlots:bsll:SR:mll:all}
\end{figure}

\begin{figure}[!h]
\centering
\subfloat[]{\includegraphics[width=0.49\textwidth]{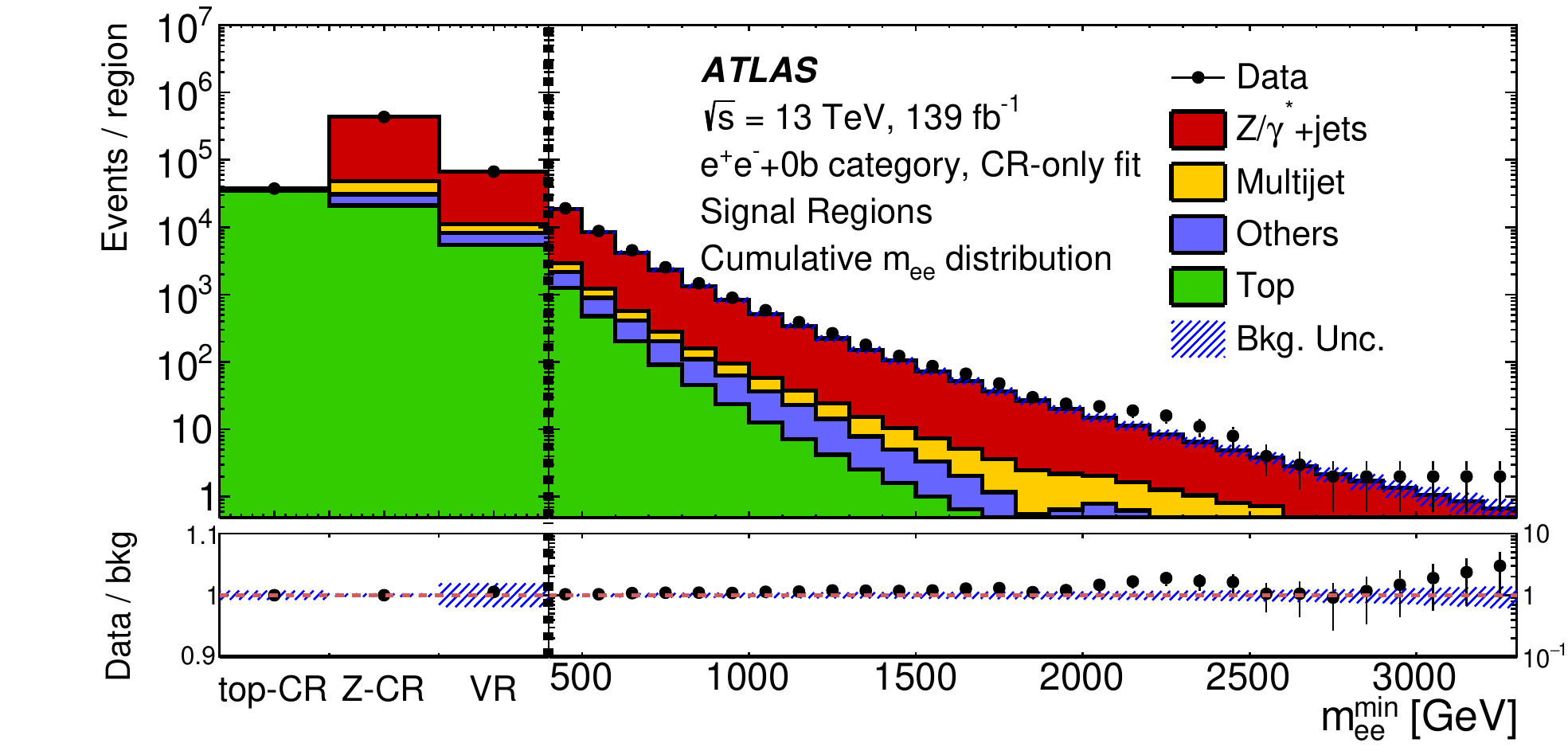}}
\subfloat[]{\includegraphics[width=0.49\textwidth]{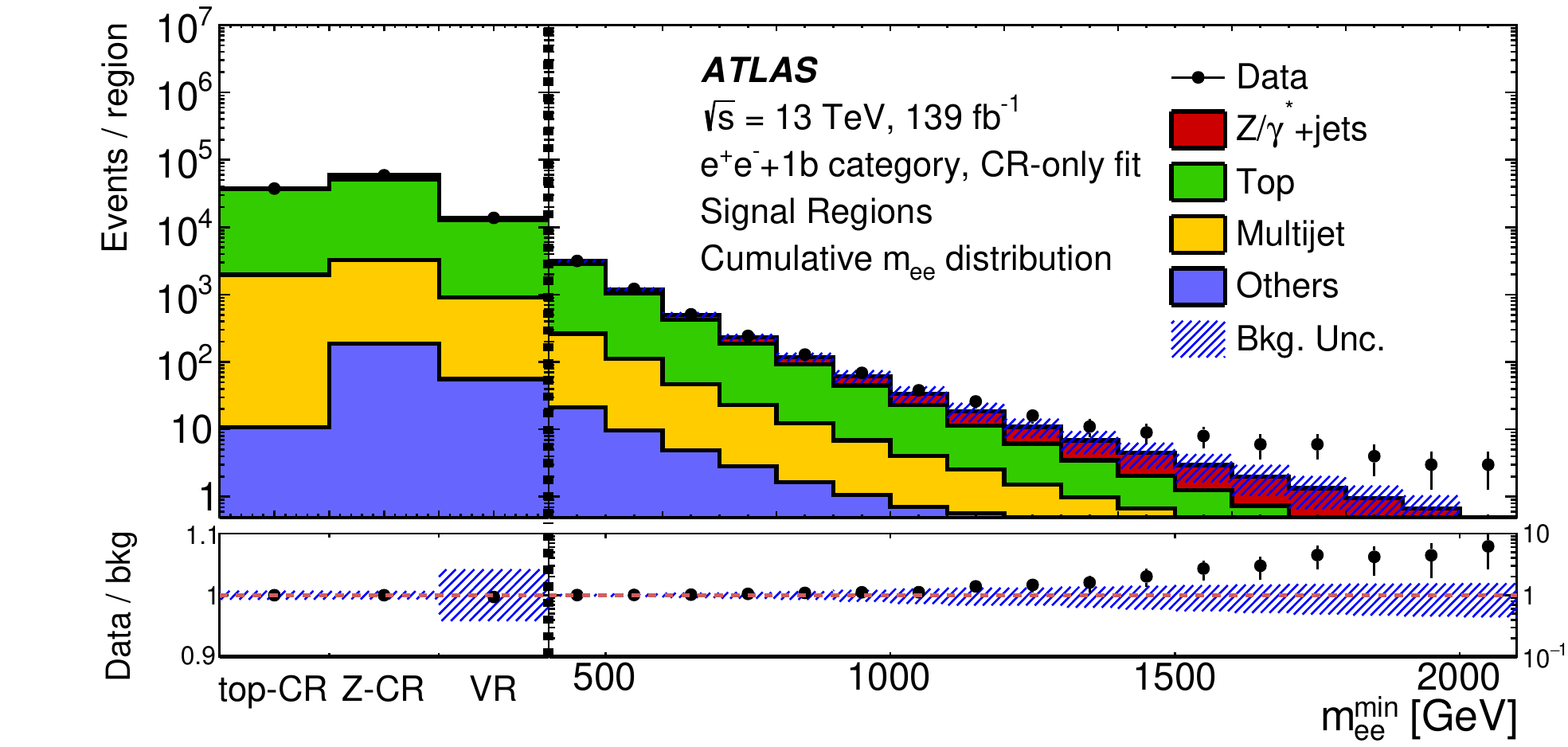}}\\
\subfloat[]{\includegraphics[width=0.49\textwidth]{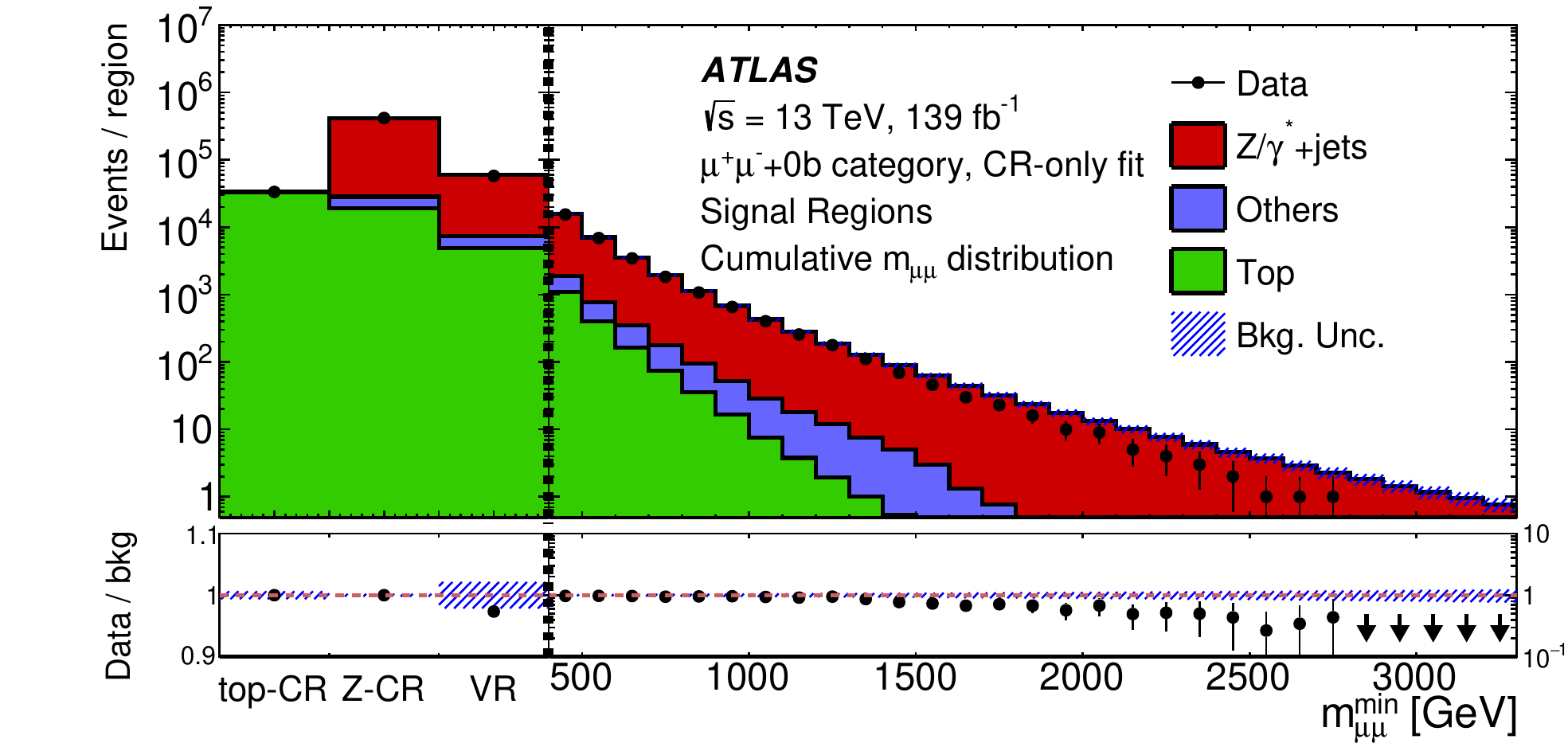}}
\subfloat[]{\includegraphics[width=0.49\textwidth]{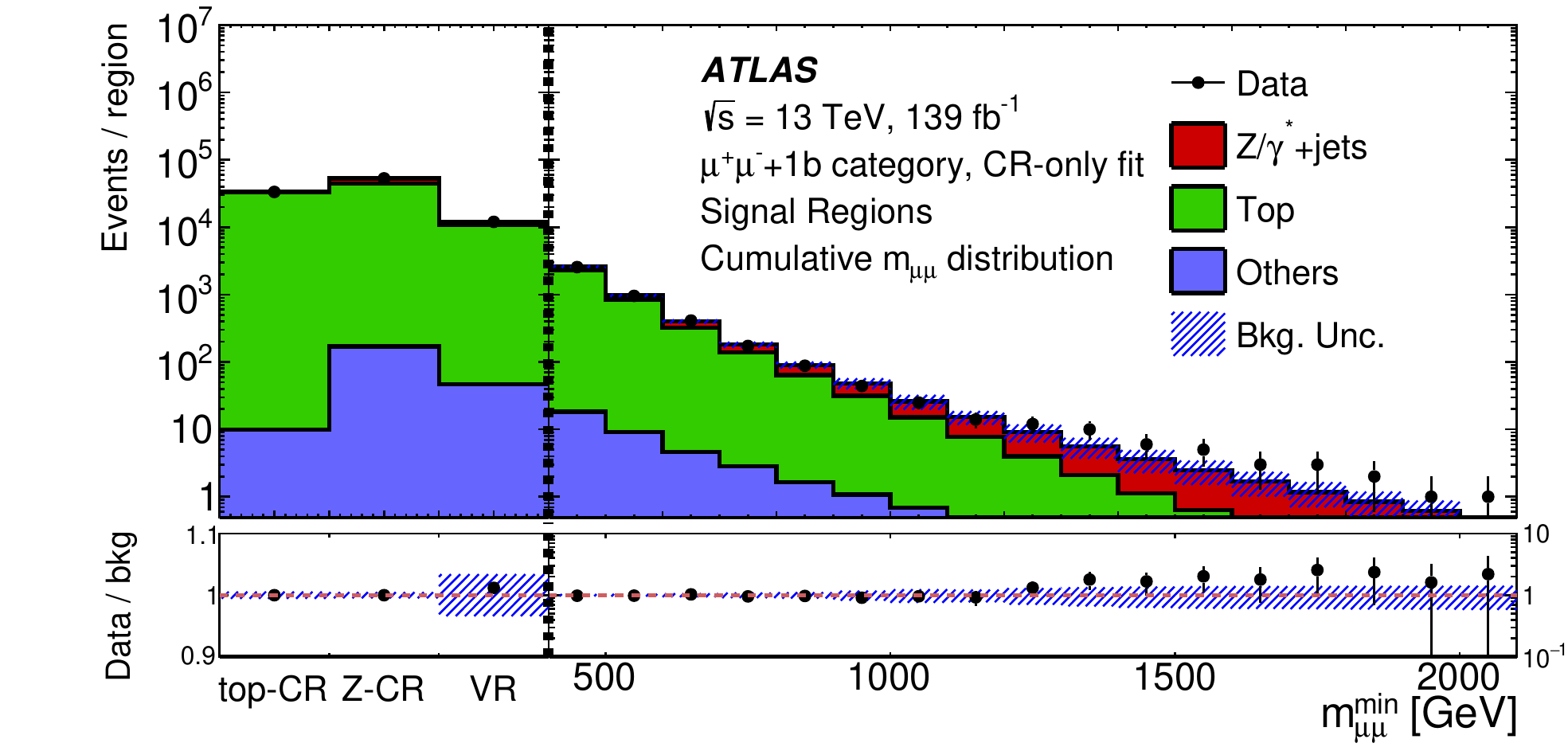}}\\
\caption{Data overlaid on SM background postfit yields in the regions of the (a) electron $b$-veto, (b) electron $b$-tag, (c) muon $b$-veto and (d) muon $b$-tag categories.
``Others'' refers to diboson and $W+$jets events.
MC statistical uncertainties and systematic uncertainties are considered (hatched band).
The left part of each figure presents the yields in the CRs and the VR of each category, while the right part presents the yields in the SRs of each category.
The bottom panels show the ratio of the data to the background prediction, while the arrows correspond to bins where the ratio is beyond the limits of the figure.
The range of the $y$-axis is different between the left and right parts of the bottom panels, and the latter is presented at logarithmic scale.
For the SRs, as the distribution is cumulative, each bin is contained in and therefore correlated with the lower mass bins.
}
\label{fig:limit:fitPlots:bsll:SR:Cum:all}
\end{figure}

\begin{figure}[!h]
\centering
\subfloat[]{\includegraphics[width=0.49\textwidth]{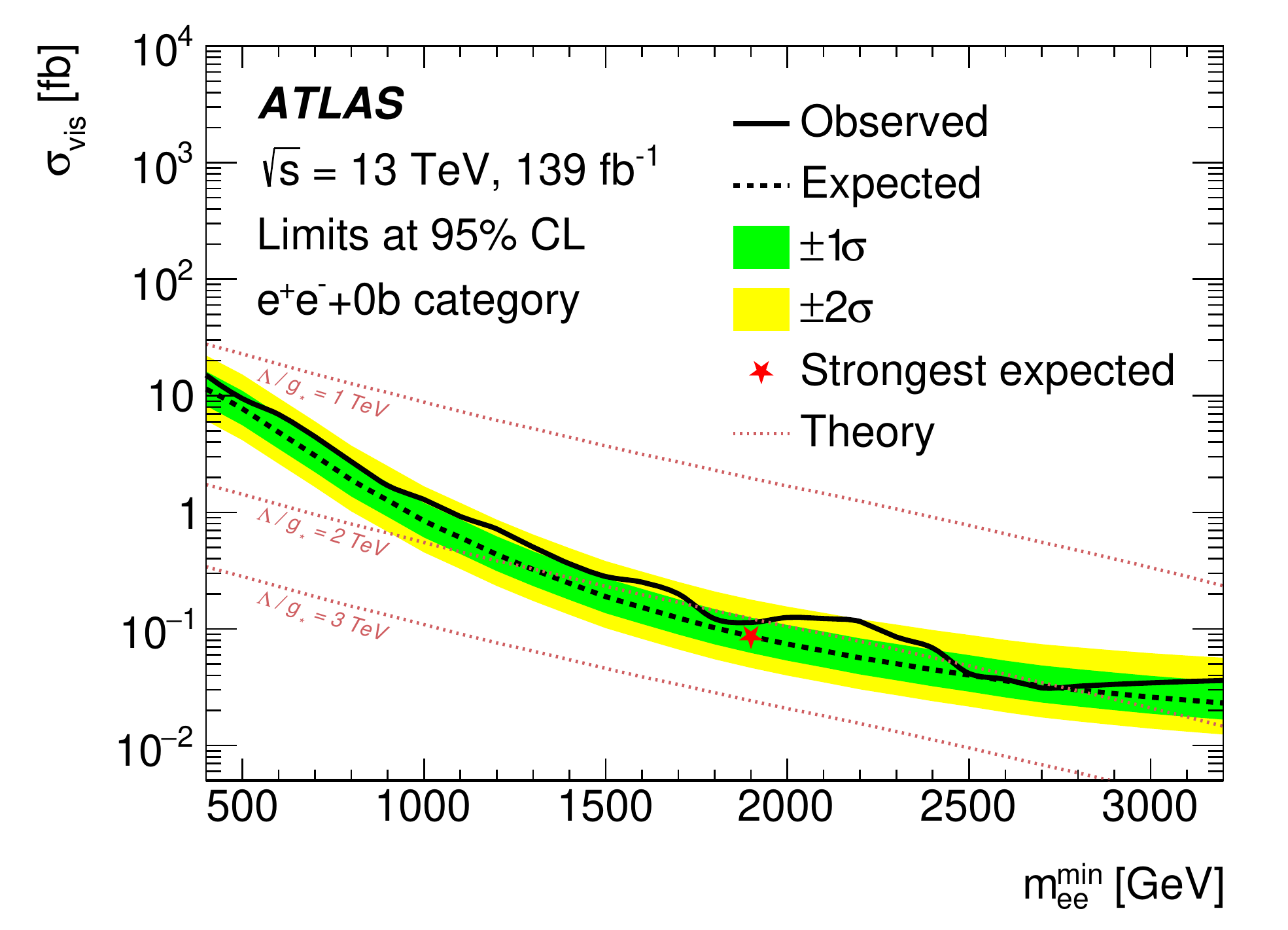}}
\subfloat[]{\includegraphics[width=0.49\textwidth]{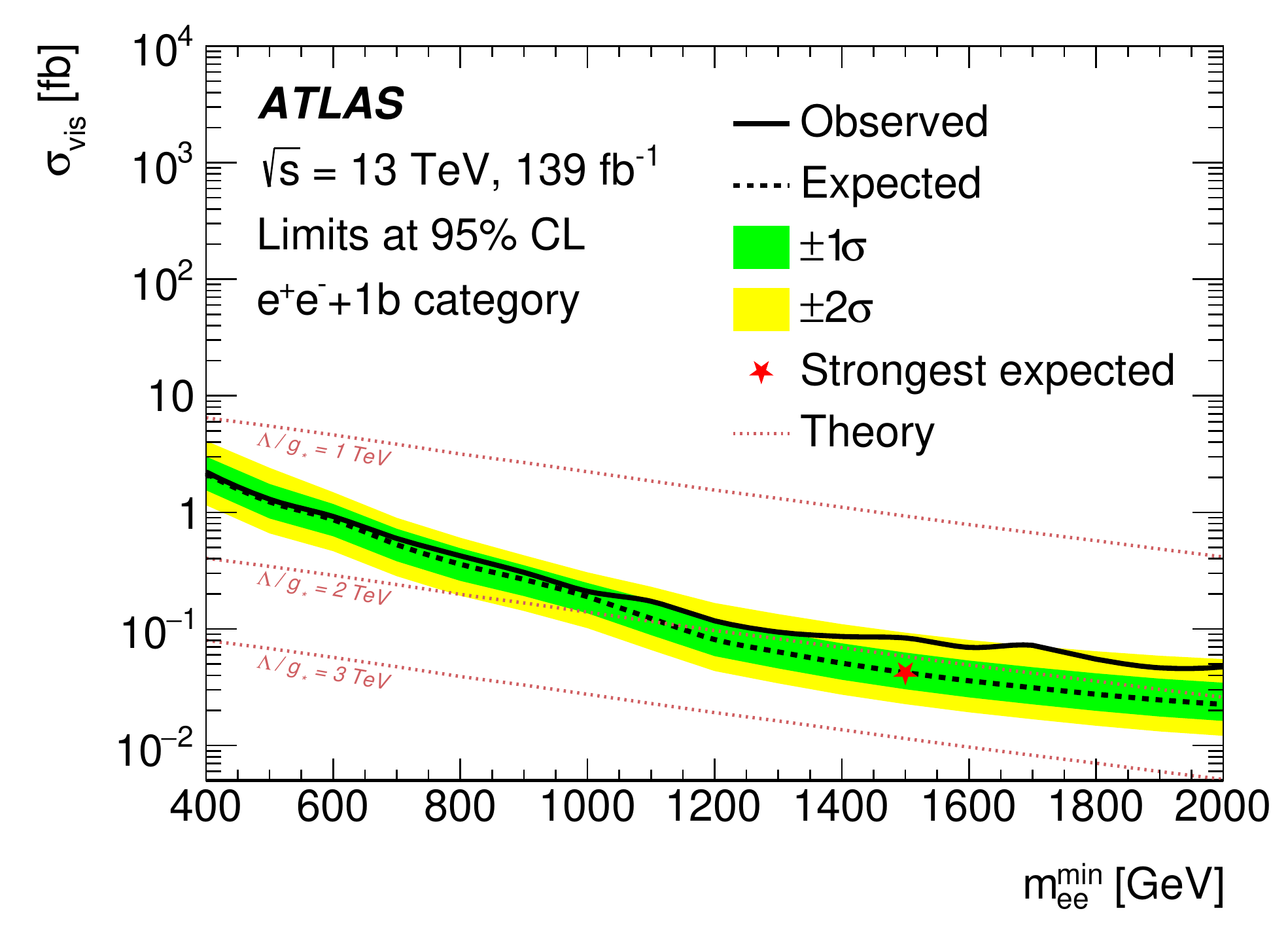}}\\
\subfloat[]{\includegraphics[width=0.49\textwidth]{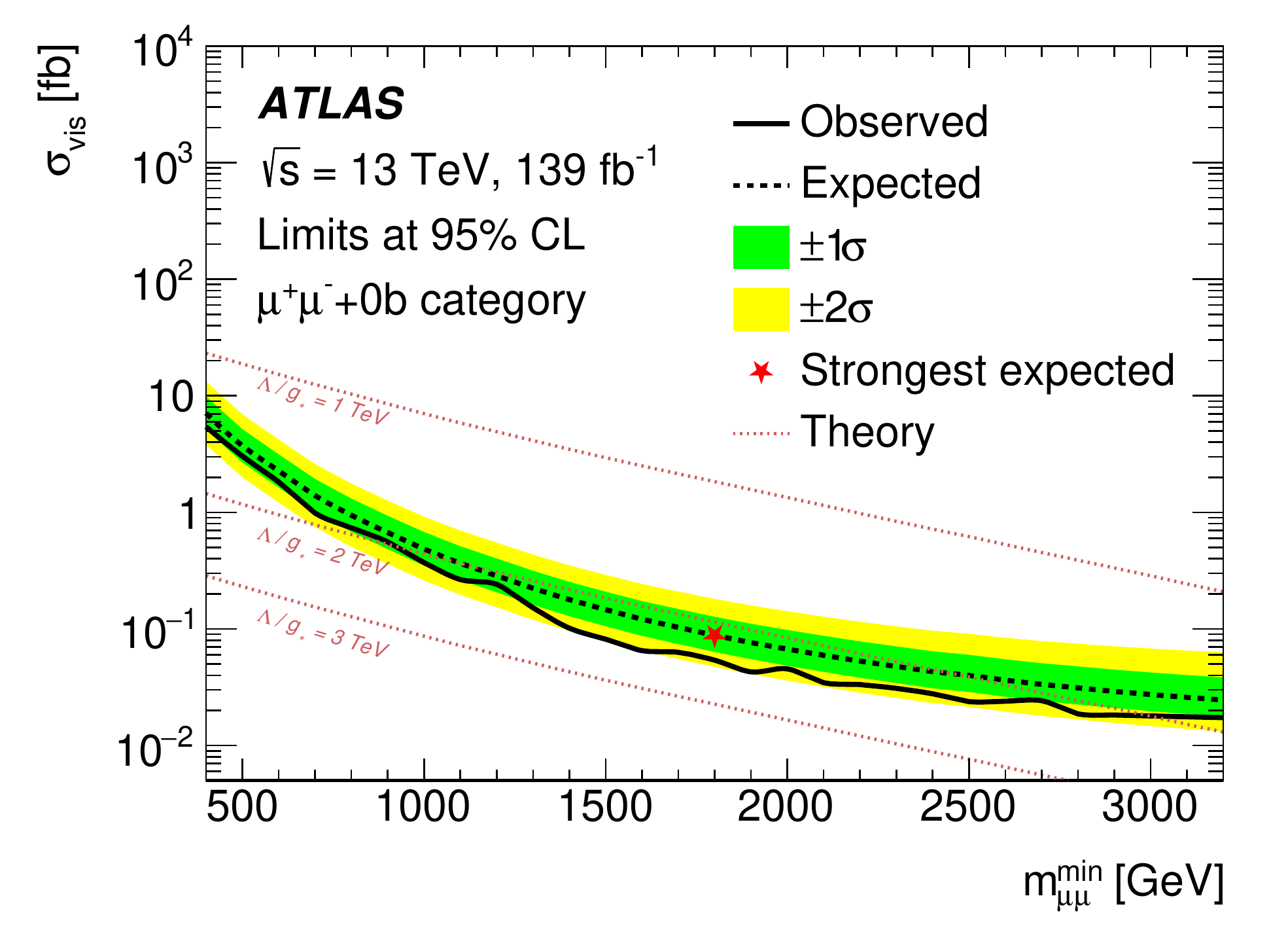}}
\subfloat[]{\includegraphics[width=0.49\textwidth]{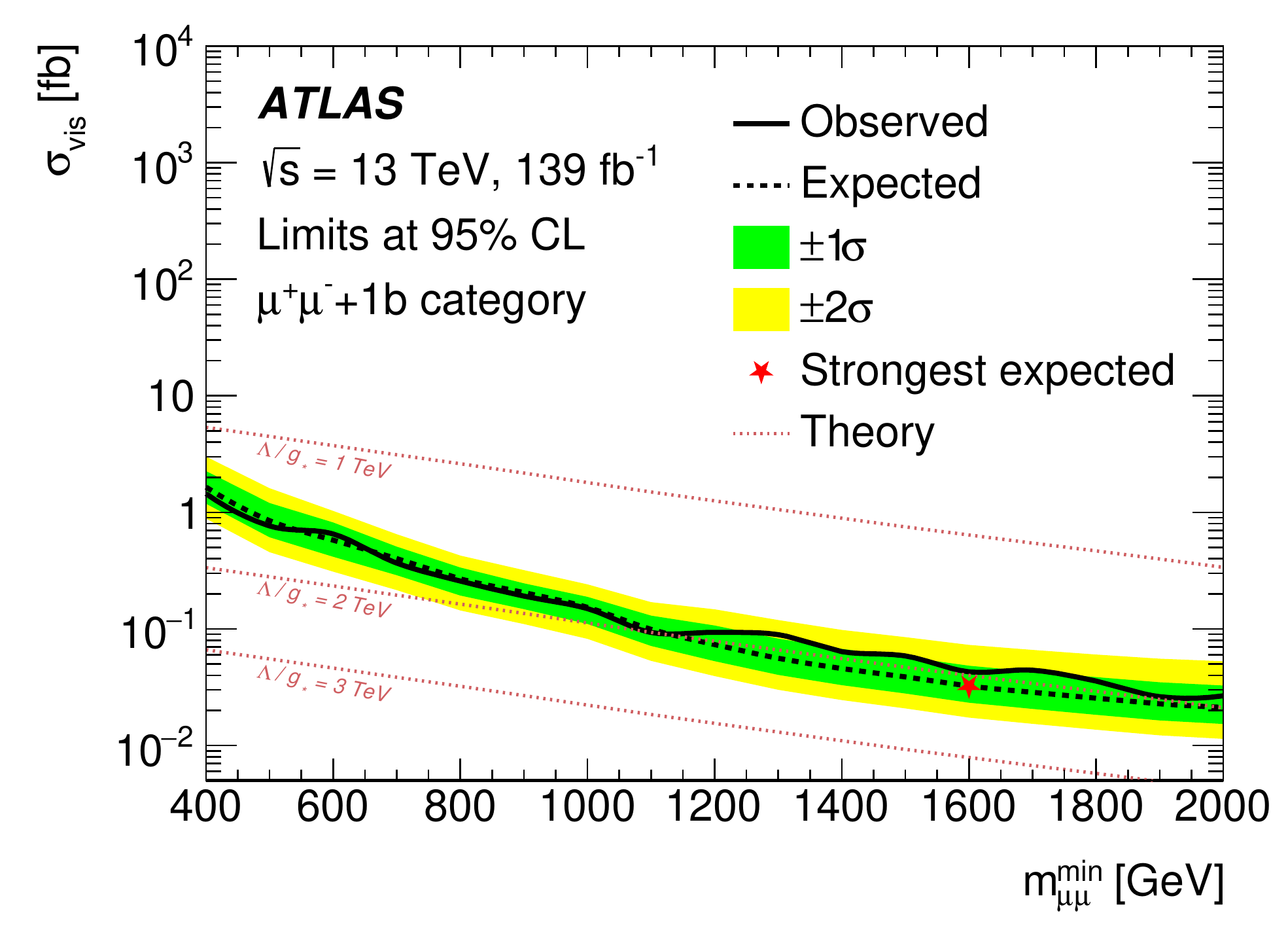}}\\
\caption{Model-independent observed (solid line) and expected (dashed line) upper limit on the visible cross section ($\sigma_{\textrm{vis}} = \sigma \cdot \epsilon \cdot \mathcal A$) for the (a) electron $b$-veto, (b) electron $b$-tag, (c) muon $b$-veto and (d) muon $b$-tag categories.
The uncertainty bands around the expected limit represent the 68\% and 95\% confidence intervals.
The theory lines (dotted lines) correspond to particular $\Lambda / g_*$ values of the signal model, and the red marker presents the strongest expected lower limit on $\Lambda / g_*$.
}
\label{fig:limit:fitPlots:bsll:SR:limits:all}
\end{figure}

\FloatBarrier
 
In summary, a search for new phenomena was conducted in final states with two electrons or muons in association with one or no $b$-tagged jets.
The analysis was conducted using 139~\ifb{} of $pp$ collision data at \sqsthirt{} recorded by the ATLAS detector at the Large Hadron Collider.
No significant excess of events above the expected SM background is observed.
Model-independent upper limits at 95\% C.L. were set on the signal cross section in each of the signal regions.
A first search for a \bsll{} contact interaction is presented, and values of $\Lambda / g_*$ smaller than 2.0 (2.4)~\TeV{} are excluded using the observed limits for electrons (muons) at 95\% CL, which is still far below the value which has been predicted in order to explain the $B$-meson decay anomalies.

\section*{Acknowledgments}

We thank CERN for the very successful operation of the LHC, as well as the
support staff from our institutions without whom ATLAS could not be
operated efficiently.
 
We acknowledge the support of ANPCyT, Argentina; YerPhI, Armenia; ARC, Australia; BMWFW and FWF, Austria; ANAS, Azerbaijan; SSTC, Belarus; CNPq and FAPESP, Brazil; NSERC, NRC and CFI, Canada; CERN; ANID, Chile; CAS, MOST and NSFC, China; Minciencias, Colombia; MSMT CR, MPO CR and VSC CR, Czech Republic; DNRF and DNSRC, Denmark; IN2P3-CNRS and CEA-DRF/IRFU, France; SRNSFG, Georgia; BMBF, HGF and MPG, Germany; GSRT, Greece; RGC and Hong Kong SAR, China; ISF and Benoziyo Center, Israel; INFN, Italy; MEXT and JSPS, Japan; CNRST, Morocco; NWO, Netherlands; RCN, Norway; MNiSW and NCN, Poland; FCT, Portugal; MNE/IFA, Romania; JINR; MES of Russia and NRC KI, Russian Federation; MESTD, Serbia; MSSR, Slovakia; ARRS and MIZ\v{S}, Slovenia; DST/NRF, South Africa; MICINN, Spain; SRC and Wallenberg Foundation, Sweden; SERI, SNSF and Cantons of Bern and Geneva, Switzerland; MOST, Taiwan; TAEK, Turkey; STFC, United Kingdom; DOE and NSF, United States of America. In addition, individual groups and members have received support from BCKDF, CANARIE, Compute Canada, CRC and IVADO, Canada; Beijing Municipal Science \& Technology Commission, China; COST, ERC, ERDF, Horizon 2020 and Marie Sk{\l}odowska-Curie Actions, European Union; Investissements d'Avenir Labex, Investissements d'Avenir Idex and ANR, France; DFG and AvH Foundation, Germany; Herakleitos, Thales and Aristeia programmes co-financed by EU-ESF and the Greek NSRF, Greece; BSF-NSF and GIF, Israel; La Caixa Banking Foundation, CERCA Programme Generalitat de Catalunya and PROMETEO and GenT Programmes Generalitat Valenciana, Spain; G\"{o}ran Gustafssons Stiftelse, Sweden; The Royal Society and Leverhulme Trust, United Kingdom.
 
The crucial computing support from all WLCG partners is acknowledged gratefully, in particular from CERN, the ATLAS Tier-1 facilities at TRIUMF (Canada), NDGF (Denmark, Norway, Sweden), CC-IN2P3 (France), KIT/GridKA (Germany), INFN-CNAF (Italy), NL-T1 (Netherlands), PIC (Spain), ASGC (Taiwan), RAL (UK) and BNL (USA), the Tier-2 facilities worldwide and large non-WLCG resource providers. Major contributors of computing resources are listed in Ref.~\cite{ATL-SOFT-PUB-2020-001}.

\clearpage
 
\begin{flushleft}
\hypersetup{urlcolor=black}
{\Large The ATLAS Collaboration}

\bigskip

\AtlasOrcid[0000-0002-6665-4934]{G.~Aad}$^\textrm{\scriptsize 99}$,    
\AtlasOrcid[0000-0002-5888-2734]{B.~Abbott}$^\textrm{\scriptsize 125}$,    
\AtlasOrcid[0000-0002-7248-3203]{D.C.~Abbott}$^\textrm{\scriptsize 100}$,    
\AtlasOrcid[0000-0002-2788-3822]{A.~Abed~Abud}$^\textrm{\scriptsize 34}$,    
\AtlasOrcid[0000-0002-1002-1652]{K.~Abeling}$^\textrm{\scriptsize 51}$,    
\AtlasOrcid[0000-0002-2987-4006]{D.K.~Abhayasinghe}$^\textrm{\scriptsize 91}$,    
\AtlasOrcid[0000-0002-8496-9294]{S.H.~Abidi}$^\textrm{\scriptsize 27}$,    
\AtlasOrcid[0000-0001-5329-6640]{H.~Abramowicz}$^\textrm{\scriptsize 158}$,    
\AtlasOrcid[0000-0002-1599-2896]{H.~Abreu}$^\textrm{\scriptsize 157}$,    
\AtlasOrcid[0000-0003-0403-3697]{Y.~Abulaiti}$^\textrm{\scriptsize 5}$
% ...
\bigskip
\\

% ...
$^{5}$High Energy Physics Division, Argonne National Laboratory, Argonne IL; United States of America.\\
% ...
$^{27}$Physics Department, Brookhaven National Laboratory, Upton NY; United States of America.\\
% ...
$^{34}$CERN, Geneva; Switzerland.\\
% ...
$^{51}$II. Physikalisches Institut, Georg-August-Universit\"{a}t G\"ottingen, G\"ottingen; Germany.\\
% ...
$^{91}$Department of Physics, Royal Holloway University of London, Egham; United Kingdom.\\
% ...
$^{99}$CPPM, Aix-Marseille Universit\'e, CNRS/IN2P3, Marseille; France.\\
$^{100}$Department of Physics, University of Massachusetts, Amherst MA; United States of America.\\
% ...
$^{125}$Homer L. Dodge Department of Physics and Astronomy, University of Oklahoma, Norman OK; United States of America.\\
% ...
$^{157}$Department of Physics, Technion, Israel Institute of Technology, Haifa; Israel.\\
$^{158}$Raymond and Beverly Sackler School of Physics and Astronomy, Tel Aviv University, Tel Aviv; Israel.\\
% ...

\end{flushleft}

 
\clearpage
\appendix

\printbibliography

@article{Martin:2004dh,
    author = "Martin, A. D. and Roberts, R. G. and Stirling, W. J. and Thorne, R. S.",
    title = "{Parton distributions incorporating QED contributions}",
    eprint = "hep-ph/0411040",
    archivePrefix = "arXiv",
    reportNumber = "IPPP-04-62, DCPT-04-124, CAVENDISH-HEP-2004-28",
    doi = "10.1140/epjc/s2004-02088-7",
    journal = "Eur. Phys. J. C",
    volume = "39",
    pages = "155--161",
    year = "2005"
}

@article{Alguero:2019ptt,
    author = "Alguer\'o, Marcel and Capdevila, Bernat and Crivellin, Andreas and Descotes-Genon, S\'ebastien and Masjuan, Pere and Matias, Joaquim and Novoa Brunet, Mart\'\i{}n and Virto, Javier",
    title = "{Emerging patterns of New Physics with and without Lepton Flavour Universal contributions}",
    eprint = "1903.09578",
    archivePrefix = "arXiv",
    primaryClass = "hep-ph",
    reportNumber = "PSI-PR-19-04, ZH-TH-14/19, TUM-HEP-1192/19, MIT-CTP/5107,
  LPT-Orsay-19-11",
    doi = "10.1140/epjc/s10052-019-7216-3",
    journal = "Eur. Phys. J. C",
    volume = "79",
    number = "8",
    pages = "714",
    year = "2019",
    note = "[Addendum: Eur. Phys. J. C 80, (2020) 511]"
}

@article{Greljo:2017vvb,
    author = "Greljo, Admir and Marzocca, David",
    title = "{High-$p_T$ dilepton tails and flavor physics}",
    eprint = "1704.09015",
    archivePrefix = "arXiv",
    primaryClass = "hep-ph",
    reportNumber = "ZU-TH-12-17",
    doi = "10.1140/epjc/s10052-017-5119-8",
    journal = "Eur. Phys. J. C",
    volume = "77",
    number = "8",
    pages = "548",
    year = "2017"
}

@article{Afik:2021jjh,
    author = "Afik, Yoav and Bar-Shalom, Shaouly and Soni, Amarjit and Wudka, Jose",
    title = "{New flavor physics in di- and tri-lepton events from single-top at the LHC and beyond}",
    eprint = "2101.05286",
    archivePrefix = "arXiv",
    primaryClass = "hep-ph",
    month = "1",
    year = "2021"
}

@article{Cacciari:2008gp,
    author = "Cacciari, Matteo and Salam, Gavin P. and Soyez, Gregory",
    title = "{The anti-$k_t$ jet clustering algorithm}",
    eprint = "0802.1189",
    archivePrefix = "arXiv",
    primaryClass = "hep-ph",
    reportNumber = "LPTHE-07-03",
    doi = "10.1088/1126-6708/2008/04/063",
    journal = "JHEP",
    volume = "04",
    pages = "063",
    year = "2008"
}

@article{Read:2002hq,
    author = "Read, Alexander L.",
    editor = "Whalley, M.R. and Lyons, L.",
    title = "{Presentation of search results: The $CL_s$ technique}",
    doi = "10.1088/0954-3899/28/10/313",
    journal = "J. Phys. G",
    volume = "28",
    pages = "2693--2704",
    year = "2002"
}

@article{Cowan:2010js,
    author = "Cowan, Glen and Cranmer, Kyle and Gross, Eilam and Vitells, Ofer",
    title = "{Asymptotic formulae for likelihood-based tests of new physics}",
    eprint = "1007.1727",
    archivePrefix = "arXiv",
    primaryClass = "physics.data-an",
    doi = "10.1140/epjc/s10052-011-1554-0",
    journal = "Eur. Phys. J. C",
    volume = "71",
    pages = "1554",
    year = "2011",
    note = "[Erratum: Eur.Phys.J.C 73, 2501 (2013)]"
}

@article{Buccioni:2019sur,
    author = {Buccioni, Federico and Lang, Jean-Nicolas and Lindert, Jonas M. and Maierh\"ofer, Philipp and Pozzorini, Stefano and Zhang, Hantian and Zoller, Max F.},
    title = "{OpenLoops 2}",
    eprint = "1907.13071",
    archivePrefix = "arXiv",
    primaryClass = "hep-ph",
    reportNumber = "IPPP/19/62, FR-PHENO-2019-12, PSI-PR-19-15, ZU-TH 37/19",
    doi = "10.1140/epjc/s10052-019-7306-2",
    journal = "Eur. Phys. J. C",
    volume = "79",
    number = "10",
    pages = "866",
    year = "2019"
}

@article{Cascioli:2011va,
    author = "Cascioli, Fabio and Maierhofer, Philipp and Pozzorini, Stefano",
    title = "{Scattering Amplitudes with Open Loops}",
    eprint = "1111.5206",
    archivePrefix = "arXiv",
    primaryClass = "hep-ph",
    reportNumber = "ZU-TH-23-11, LPN11-66",
    doi = "10.1103/PhysRevLett.108.111601",
    journal = "Phys. Rev. Lett.",
    volume = "108",
    pages = "111601",
    year = "2012"
}

@article{Denner:2016kdg,
    author = "Denner, Ansgar and Dittmaier, Stefan and Hofer, Lars",
    title = "{Collier: A fortran-based complex one-loop lIbrary in extended regularizations}",
    eprint = "1604.06792",
    archivePrefix = "arXiv",
    primaryClass = "hep-ph",
    reportNumber = "FR-PHENO-2016-003, ICCUB-16-016",
    doi = "10.1016/j.cpc.2016.10.013",
    journal = "Comput. Phys. Commun.",
    volume = "212",
    pages = "220--238",
    year = "2017"
}

@article{Hoeche:2011fd,
    author = "Hoeche, Stefan and Krauss, Frank and Schonherr, Marek and Siegert, Frank",
    title = "{A critical appraisal of NLO+PS matching methods}",
    eprint = "1111.1220",
    archivePrefix = "arXiv",
    primaryClass = "hep-ph",
    reportNumber = "SLAC-PUB-14661, IPPP-11-67, DCPT-11-134, LPN11-58, FR-PHENO-2011-019, MCNET-11-24",
    doi = "10.1007/JHEP09(2012)049",
    journal = "JHEP",
    volume = "09",
    pages = "049",
    year = "2012"
}

@article{Catani:2001cc,
    author = "Catani, S. and Krauss, F. and Kuhn, R. and Webber, B.R.",
    title = "{QCD Matrix Elements + Parton Showers}",
    eprint = "hep-ph/0109231",
    archivePrefix = "arXiv",
    reportNumber = "CERN-TH-2000-367, CAVENDISH-HEP-00-03",
    doi = "10.1088/1126-6708/2001/11/063",
    journal = "JHEP",
    volume = "11",
    pages = "063",
    year = "2001"
}

@article{Bothmann:2019yzt,
      author         = "Bothmann, Enrico and others",
      title          = "{Event generation with Sherpa 2.2}",
      year           = "2019",
      eprint         = "1905.09127",
      archivePrefix  = "arXiv",
      primaryClass   = "hep-ph",
      reportNumber   = "FERMILAB-PUB-19-218-T, SLAC-PUB-17433, IPPP/19/42,
                        MCNET-19-11",
journal={SciPost Phys.},
	volume={7},
	pages={034},
	year={2019},
	publisher={SciPost},
	doi={10.21468/SciPostPhys.7.3.034},			
      SLACcitation   = "%%CITATION = ARXIV:1905.09127;%%"
}

@article{Agostinelli:2002hh,
      author         = "Agostinelli, S. and others",
      title          = "{Geant4--a simulation toolkit}",
      collaboration  = "GEANT4",
      journal        = "Nucl. Instrum. Meth. A",
      volume         = "506",
      year           = "2003",
      pages          = "250",
      doi            = "10.1016/S0168-9002(03)01368-8",
      reportNumber   = "SLAC-PUB-9350, FERMILAB-PUB-03-339",
      SLACcitation   = "%%CITATION = NUIMA,A506,250;%%"
}

@Article{Sjostrand:2007gs,
      author         = "Sj{\"o}strand, Torbjorn and Mrenna, Stephen and Skands, Peter
                        Z.",
      title          = "{A brief introduction to PYTHIA 8.1}",
      journal        = "Comput. Phys. Commun.",
      volume         = "178",
      pages          = "852",
      doi            = "10.1016/j.cpc.2008.01.036",
      year           = "2008",
      eprint         = "0710.3820",
      archivePrefix  = "arXiv",
      primaryClass   = "hep-ph",
      reportNumber   = "CERN-LCGAPP-2007-04, LU-TP-07-28,
                        FERMILAB-PUB-07-512-CD-T",
      SLACcitation   = "%%CITATION = ARXIV:0710.3820;%%",
}

@Article{Alwall:2014hca,
      author         = "Alwall, J. and Frederix, R. and Frixione, S. and Hirschi,
                        V. and Maltoni, F. and others",
      title          = "{The automated computation of tree-level and
                        next-to-leading order differential cross sections, and
                        their matching to parton shower simulations}",
      journal        = "JHEP",
      volume         = "07",
      pages          = "079",
      doi            = "10.1007/JHEP07(2014)079",
      year           = "2014",
      eprint         = "1405.0301",
      archivePrefix  = "arXiv",
      primaryClass   = "hep-ph"
}

@Article{Nason:2004rx,
      author         = "Nason, Paolo",
      title          = "{A new method for combining NLO QCD with shower Monte
                        Carlo algorithms}",
      journal        = "JHEP",
      volume         = "11",
      pages          = "040",
      doi            = "10.1088/1126-6708/2004/11/040",
      year           = "2004",
      eprint         = "hep-ph/0409146",
      archivePrefix  = "arXiv",
}

@Article{Frixione:2007vw,
      author         = "Frixione, Stefano and Nason, Paolo and Oleari, Carlo",
      title          = "{Matching NLO QCD computations with parton shower
                        simulations: the POWHEG method}",
      journal        = "JHEP",
      volume         = "11",
      pages          = "070",
      doi            = "10.1088/1126-6708/2007/11/070",
      year           = "2007",
      eprint         = "0709.2092",
      archivePrefix  = "arXiv",
      primaryClass   = "hep-ph",
}

@article{Ball:2014uwa,
      author         = "{NNPDF Collaboration, R.D. Ball et al.}",
      title          = "{Parton distributions for the LHC run II}",
      collaboration  = "NNPDF",
      journal        = "JHEP",
      volume         = "04",
      pages          = "040",
      doi            = "10.1007/JHEP04(2015)040",
      year           = "2015",
      eprint         = "1410.8849",
      archivePrefix  = "arXiv",
      primaryClass   = "hep-ph",
      reportNumber   = "EDINBURGH-2014-15, IFUM-1034-FT, CERN-PH-TH-2013-253,
                        OUTP-14-11P, CAVENDISH-HEP-14-11",
      SLACcitation   = "%%CITATION = ARXIV:1410.8849;%%",
}

@article{Sjostrand:2014zea,
      author         = "Sj{\"o}strand, Torbj{\"o}rn and Ask, Stefan and Christiansen,
                        Jesper R. and Corke, Richard and Desai, Nishita and Ilten,
                        Philip and Mrenna, Stephen and Prestel, Stefan and
                        Rasmussen, Christine O. and Skands, Peter Z.",
      title          = "{An introduction to PYTHIA 8.2}",
      journal        = "Comput. Phys. Commun.",
      volume         = "191",
      year           = "2015",
      pages          = "159-177",
      doi            = "10.1016/j.cpc.2015.01.024",
      eprint         = "1410.3012",
      archivePrefix  = "arXiv",
      primaryClass   = "hep-ph",
      reportNumber   = "LU-TP-14-36, MCNET-14-22, CERN-PH-TH-2014-190,
                        FERMILAB-PUB-14-316-CD, DESY-14-178, SLAC-PUB-16122,
                        --FERMILAB-PUB-14-316-CD",
      SLACcitation   = "%%CITATION = ARXIV:1410.3012;%%"
}

@article{Frixione:2008yi,
      author         = "Frixione, Stefano and Laenen, Eric and Motylinski,
                        Patrick and Webber, Bryan R. and White, Chris D.",
      title          = "{Single-top hadroproduction in association with a W
                        boson}",
      journal        = "JHEP",
      volume         = "07",
      year           = "2008",
      pages          = "029",
      doi            = "10.1088/1126-6708/2008/07/029",
      eprint         = "0805.3067",
      archivePrefix  = "arXiv",
      primaryClass   = "hep-ph",
      reportNumber   = "CERN-TH-2008-104, ITP-UU-08-27, NIKHEF-2008-005,
                        ITFA-2008-16, CAVENDISH-HEP-08-06",
      SLACcitation   = "%%CITATION = ARXIV:0805.3067;%%"
}

@Article{Alioli:2010xd,
      author         = "Alioli, Simone and Nason, Paolo and Oleari, Carlo and Re,
                        Emanuele",
      title          = "{A general framework for implementing NLO calculations in
                        shower Monte Carlo programs: the POWHEG BOX}",
      journal        = "JHEP",
      volume         = "06",
      pages          = "043",
      doi            = "10.1007/JHEP06(2010)043",
      year           = "2010",
      eprint         = "1002.2581",
      archivePrefix  = "arXiv",
      primaryClass   = "hep-ph",
}

@Article{Hoeche:2009rj,
      author         = "H{\"o}che, Stefan and Krauss, Frank and Schumann, Steffen
                        and Siegert, Frank",
      title          = "{QCD matrix elements and truncated showers}",
      journal        = "JHEP",
      volume         = "05",
      pages          = "053",
      doi            = "10.1088/1126-6708/2009/05/053",
      year           = "2009",
      eprint         = "0903.1219",
      archivePrefix  = "arXiv",
      primaryClass   = "hep-ph",
}

@Article{Gleisberg:2008fv,
      author         = "Gleisberg, Tanju and H{\"o}che, Stefan",
      title          = "{Comix, a new matrix element generator}",
      journal        = "JHEP",
      volume         = "12",
      pages          = "039",
      doi            = "10.1088/1126-6708/2008/12/039",
      year           = "2008",
      eprint         = "0808.3674",
      archivePrefix  = "arXiv",
      primaryClass   = "hep-ph",
}

@Article{Schumann:2007mg,
      author         = "Schumann, Steffen and Krauss, Frank",
      title          = "{A parton shower algorithm based on Catani-Seymour dipole
                        factorisation}",
      journal        = "JHEP",
      volume         = "03",
      pages          = "038",
      doi            = "10.1088/1126-6708/2008/03/038",
      year           = "2008",
      eprint         = "0709.1027",
      archivePrefix  = "arXiv",
      primaryClass   = "hep-ph",
}

@Article{Lange:2001uf,
      author         = "Lange, D. J.",
      title          = "{The EvtGen particle decay simulation package}",
      booktitle      = "{Proceedings, 7th International Conference on B physics
                        at hadron machines (BEAUTY 2000)}",
      journal        = "Nucl. Instrum. Meth. A",
      volume         = "462",
      year           = "2001",
      pages          = "152",
      doi            = "10.1016/S0168-9002(01)00089-4",
      SLACcitation   = "%%CITATION = NUIMA,A462,152;%%"
}

@article{ALEPH:2005ab,
      author         = "Schael, S. and others",
      title          = "{Precision electroweak measurements on the $Z$
                        resonance}",
      collaboration  = "SLD Electroweak Group, DELPHI, ALEPH, SLD, SLD Heavy
                        Flavour Group, OPAL, LEP Electroweak Working Group, L3",
      journal        = "Phys. Rept.",
      volume         = "427",
      year           = "2006",
      pages          = "257-454",
      doi            = "10.1016/j.physrep.2005.12.006",
      eprint         = "hep-ex/0509008",
      archivePrefix  = "arXiv",
      primaryClass   = "hep-ex",
      reportNumber   = "SLAC-R-774",
      SLACcitation   = "%%CITATION = HEP-EX/0509008;%%"
}

@article{Afik:2018nlr,
      author         = "Afik, Yoav and Cohen, Jonathan and Gozani, Eitan and
                        Kajomovitz, Enrique and Rozen, Yoram",
      title          = "{Establishing a search for $b \rightarrow s \ell^{+} \ell^{-}$ anomalies at the LHC}",
      journal        = "JHEP",
      volume         = "08",
      year           = "2018",
      pages          = "056",
      doi            = "10.1007/JHEP08(2018)056",
      eprint         = "1805.11402",
      archivePrefix  = "arXiv",
      primaryClass   = "hep-ph",
      SLACcitation   = "%%CITATION = ARXIV:1805.11402;%%"
}

@article{Afik:2019htr,
    author = "Afik, Yoav and Bar-Shalom, Shaouly and Cohen, Jonathan and Rozen, Yoram",
    title = "{Searching for New Physics with $b\bar{b} \ell^+ \ell^-$ contact interactions}",
    eprint = "1912.00425",
    archivePrefix = "arXiv",
    primaryClass = "hep-ex",
    doi = "10.1016/j.physletb.2020.135541",
    journal = "Phys. Lett. B",
    volume = "807",
    pages = "135541",
    year = "2020"
}

@article{Alioli:2008gx,
      author         = "Alioli, Simone and Nason, Paolo and Oleari, Carlo and Re,
                        Emanuele",
      title          = "{NLO vector-boson production matched with shower in
                        POWHEG}",
      journal        = "JHEP",
      volume         = "07",
      year           = "2008",
      pages          = "060",
      doi            = "10.1088/1126-6708/2008/07/060",
      eprint         = "0805.4802",
      archivePrefix  = "arXiv",
      primaryClass   = "hep-ph",
      reportNumber   = "NSF-KITP-08-74",
      SLACcitation   = "%%CITATION = ARXIV:0805.4802;%%"
}

@Article{Pumplin:2002vw,
     author    = "Pumplin, J. and others",
     title     = "{New Generation of Parton Distributions with Uncertainties
                  From Global QCD Analysis}",
     journal   = "JHEP",
     volume    = "07",
     year      = "2002",
     pages     = "012",
     doi       = "10.1088/1126-6708/2002/07/012",
     eprint    = "hep-ph/0201195",
     archivePrefix = "arXiv",
     SLACcitation  = "%%CITATION = HEP-PH/0201195;%%"
}

@article{Lai:2010vv,
      author         = "Lai, Hung-Liang and Guzzi, Marco and Huston, Joey and Li,
                        Zhao and Nadolsky, Pavel M. and others",
      title          = "{New parton distributions for collider physics}",
      journal        = "Phys. Rev. D",
      volume         = "82",
      pages          = "074024",
      doi            = "10.1103/PhysRevD.82.074024",
      year           = "2010",
      eprint         = "1007.2241",
      archivePrefix  = "arXiv",
      primaryClass   = "hep-ph",
      reportNumber   = "MSUHEP-100707, SMU-HEP-10-10",
      SLACcitation   = "%%CITATION = ARXIV:1007.2241;%%",
}

@article{Golonka:2005pn,
      author         = "Golonka, Piotr and Was, Zbigniew",
      title          = "{PHOTOS Monte Carlo: a precision tool for QED corrections
                        in $Z$ and $W$ decays}",
      journal        = "Eur. Phys. J. C",
      volume         = "45",
      year           = "2006",
      pages          = "97-107",
      doi            = "10.1140/epjc/s2005-02396-4",
      eprint         = "hep-ph/0506026",
      archivePrefix  = "arXiv",
      primaryClass   = "hep-ph",
      reportNumber   = "IFJPAN-V-05-01, CERN-PH-TH-2005-091",
      SLACcitation   = "%%CITATION = HEP-PH/0506026;%%"
}

@article{Davidson:2010ew,
      author         = "Davidson, N. and Przedzinski, T. and Was, Z.",
      title          = "{PHOTOS interface in C++: Technical and physics
                        documentation}",
      journal        = "Comput. Phys. Commun.",
      volume         = "199",
      year           = "2016",
      pages          = "86-101",
      doi            = "10.1016/j.cpc.2015.09.013",
      eprint         = "1011.0937",
      archivePrefix  = "arXiv",
      primaryClass   = "hep-ph",
      reportNumber   = "CERN-PH-TH-2010-261, IFJPAN-IV-2010-6",
      SLACcitation   = "%%CITATION = ARXIV:1011.0937;%%"
}

@article{Bellm:2015jjp,
    author = "Bellm, Johannes and others",
    title = "{Herwig 7.0/Herwig++ 3.0 release note}",
    eprint = "1512.01178",
    archivePrefix = "arXiv",
    primaryClass = "hep-ph",
    reportNumber = "CERN-PH-TH-2015-289, MAN-HEP-2015-15, IFJPAN-IV-2015-13, KA-TP-18-2015, DCPT-15-142, MCNET-15-28, IPPP-15-71, HERWIG-2015-01",
    doi = "10.1140/epjc/s10052-016-4018-8",
    journal = "Eur. Phys. J. C",
    volume = "76",
    number = "4",
    pages = "196",
    year = "2016"
}

@article{Gao:2013bia,
      author         = "Gao, Jun and Nadolsky, Pavel",
      title          = "{A meta-analysis of parton distribution functions}",
      journal        = "JHEP",
      volume         = "07",
      year           = "2014",
      pages          = "035",
      doi            = "10.1007/JHEP07(2014)035",
      eprint         = "1401.0013",
      archivePrefix  = "arXiv",
      primaryClass   = "hep-ph",
      SLACcitation   = "%%CITATION = ARXIV:1401.0013;%%"
}

@article{Butterworth:2015oua,
    author = "Butterworth, Jon and others",
    title = "{PDF4LHC recommendations for LHC Run II}",
    eprint = "1510.03865",
    archivePrefix = "arXiv",
    primaryClass = "hep-ph",
    reportNumber = "OUTP-15-17P, SMU-HEP-15-12, TIF-UNIMI-2015-14, LCTS-2015-27, CERN-PH-TH-2015-249",
    doi = "10.1088/0954-3899/43/2/023001",
    journal = "J. Phys. G",
    volume = "43",
    pages = "023001",
    year = "2016"
}

@article{Anastasiou:2003ds,
    author = "Anastasiou, Charalampos and Dixon, Lance J. and Melnikov, Kirill and Petriello, Frank",
    title = "{High precision QCD at hadron colliders: Electroweak gauge boson rapidity distributions at NNLO}",
    eprint = "hep-ph/0312266",
    archivePrefix = "arXiv",
    reportNumber = "SLAC-PUB-10288, UH-511-1042-03",
    doi = "10.1103/PhysRevD.69.094008",
    journal = "Phys. Rev. D",
    volume = "69",
    pages = "094008",
    year = "2004"
}

@article{Lonnblad:2011xx,
      author         = "Lönnblad, Leif and Prestel, Stefan",
      title          = "{Matching tree-level matrix elements with interleaved showers}",
      journal        = "JHEP",
      volume         = "03",
      year           = "2012",
      pages          = "019",
      doi            = "10.1007/JHEP03(2012)019",
      eprint         = "1109.4829",
      archivePrefix  = "arXiv",
      primaryClass   = "hep-ph",
      SLACcitation   = "%%CITATION = ARXIV:1109.4829;%%"
}

@article{Hoeche:2012yf,
      author         = "Hoeche, Stefan and Krauss, Frank and Schonherr, Marek and
                        Siegert, Frank",
      title          = "{QCD matrix elements + parton showers. The NLO case}",
      journal        = "JHEP",
      volume         = "04",
      year           = "2013",
      pages          = "027",
      doi            = "10.1007/JHEP04(2013)027",
      eprint         = "1207.5030",
      archivePrefix  = "arXiv",
      primaryClass   = "hep-ph",
      reportNumber   = "SLAC-PUB-15191, IPPP-12-52, DCPT-12-104, LPN12-081,
                        MCNET-12-09, FR-PHENO-2012-017",
      SLACcitation   = "%%CITATION = ARXIV:1207.5030;%%"
}

@article{Aaij:2014pli,
     author         = "{LHCb Collaboration}",
    collaboration = "LHCb",
    title = "{Differential branching fractions and isospin asymmetries of $B \to K^{(*)} \mu^+ \mu^-$ decays}",
    eprint = "1403.8044",
    archivePrefix = "arXiv",
    primaryClass = "hep-ex",
    reportNumber = "LHCB-PAPER-2014-006, CERN-PH-EP-2014-055",
    doi = "10.1007/JHEP06(2014)133",
    journal = "JHEP",
    volume = "06",
    pages = "133",
    year = "2014"
}

@article{Aaij:2015esa,
    author         = "{LHCb Collaboration}",
    collaboration = "LHCb",
    title = "{Angular analysis and differential branching fraction of the decay $B^0_s\to\phi\mu^+\mu^-$}",
    eprint = "1506.08777",
    archivePrefix = "arXiv",
    primaryClass = "hep-ex",
    reportNumber = "CERN-PH-EP-2015-145, LHCB-PAPER-2015-023",
    doi = "10.1007/JHEP09(2015)179",
    journal = "JHEP",
    volume = "09",
    pages = "179",
    year = "2015"
}

@article{Aaij:2015xza,
    author         = "{LHCb Collaboration}",
    collaboration = "LHCb",
    title = "{Differential branching fraction and angular analysis of $\Lambda^{0}_{b} \rightarrow \Lambda \mu^+\mu^-$ decays}",
    eprint = "1503.07138",
    archivePrefix = "arXiv",
    primaryClass = "hep-ex",
    reportNumber = "LHCB-PAPER-2015-009, CERN-PH-EP-2015-078",
    doi = "10.1007/JHEP06(2015)115",
    journal = "JHEP",
    volume = "06",
    pages = "115",
    year = "2015",
    note = "[Erratum: JHEP 09, 145 (2018)]"
}

@article{Aaij:2016flj,
    author         = "{LHCb Collaboration}",
    collaboration = "LHCb",
    title = "{Measurements of the S-wave fraction in $B^{0}\rightarrow K^{+}\pi^{-}\mu^{+}\mu^{-}$ decays and the $B^{0}\rightarrow K^{\ast}(892)^{0}\mu^{+}\mu^{-}$ differential branching fraction}",
    eprint = "1606.04731",
    archivePrefix = "arXiv",
    primaryClass = "hep-ex",
    reportNumber = "CERN-EP-2016-141, LHCB-PAPER-2016-012",
    doi = "10.1007/JHEP11(2016)047",
    journal = "JHEP",
    volume = "11",
    pages = "047",
    year = "2016",
    note = "[Erratum: JHEP 04, (2017) 142]"
}

@article{Aaij:2017vbb,
    author         = "{LHCb Collaboration}",
    collaboration = "LHCb",
    title = "{Test of lepton universality with $B^{0} \rightarrow K^{*0}\ell^{+}\ell^{-}$ decays}",
    eprint = "1705.05802",
    archivePrefix = "arXiv",
    primaryClass = "hep-ex",
    reportNumber = "LHCB-PAPER-2017-013, CERN-EP-2017-100",
    doi = "10.1007/JHEP08(2017)055",
    journal = "JHEP",
    volume = "08",
    pages = "055",
    year = "2017"
}

@article{Aaij:2019wad,
    author         = "{LHCb Collaboration}",
    collaboration = "LHCb",
    title = "{Search for lepton-universality violation in $B^+\to K^+\ell^+\ell^-$ decays}",
    eprint = "1903.09252",
    archivePrefix = "arXiv",
    primaryClass = "hep-ex",
    reportNumber = "LHCb-PAPER-2019-009, CERN-EP-2019-043, LHCb-PAPER-2019-009 CERN-EP-2019-043",
    doi = "10.1103/PhysRevLett.122.191801",
    journal = "Phys. Rev. Lett.",
    volume = "122",
    number = "19",
    pages = "191801",
    year = "2019"
}

@article{Aaij:2019bzx,
    author         = "{LHCb Collaboration}",
    collaboration = "LHCb",
    title = "{Test of lepton universality with $ {\Lambda}_b^0\to {pK}^{-}{\mathrm{\ell}}^{+}{\mathrm{\ell}}^{-} $ decays}",
    eprint = "1912.08139",
    archivePrefix = "arXiv",
    primaryClass = "hep-ex",
    reportNumber = "LHCb-PAPER-2019-040, CERN-EP-2019-272, LHCB-PAPER-2019-040, LHCB-PAPER-2019-040-003",
    doi = "10.1007/JHEP05(2020)040",
    journal = "JHEP",
    volume = "05",
    pages = "040",
    year = "2020"
}

@article{Lees:2012tva,
    author = "{BaBar Collaboration}",
    collaboration = "BaBar",
    title = "{Measurement of branching fractions and rate asymmetries in the rare decays $B\ensuremath{\rightarrow}{K}^{(*)}{\ensuremath{\ell}}^{+}{\ensuremath{\ell}}^{\ensuremath{-}}$}",
    eprint = "1204.3933",
    archivePrefix = "arXiv",
    primaryClass = "hep-ex",
    reportNumber = "SLAC-PUB-14957, BABAR-PUB-12-002",
    doi = "10.1103/PhysRevD.86.032012",
    journal = "Phys. Rev. D",
    volume = "86",
    pages = "032012",
    year = "2012"
}

@article{Abdesselam:2019wac,
    author = "{Belle Collaboration}",
    collaboration = "Belle",
    title = "{Test of lepton-flavor universality in ${B\to K^\ast\ell^+\ell^-}$ decays at Belle}",
    eprint = "1904.02440",
    archivePrefix = "arXiv",
    primaryClass = "hep-ex",
    reportNumber = "BELLE-CONF-1901",
    month = "4",
    year = "2019"
}

@article{Abdesselam:2019lab,
    author = "Choudhury, S. and others",
    collaboration = "BELLE",
    title = "{Test of lepton flavor universality and search for lepton flavor violation in $B \rightarrow K\ell \ell$ decays}",
    eprint = "1908.01848",
    archivePrefix = "arXiv",
    primaryClass = "hep-ex",
    reportNumber = "BELLE-CONF-1904",
    doi = "10.1007/JHEP03(2021)105",
    journal = "JHEP",
    volume = "03",
    pages = "105",
    year = "2021"
}

@article{Aaij:2015oid,
    author         = "{LHCb Collaboration}",
    collaboration = "LHCb",
    title = "{Angular analysis of the $B^{0} \to K^{*0} \mu^{+} \mu^{-}$ decay using 3 fb$^{-1}$ of integrated luminosity}",
    eprint = "1512.04442",
    archivePrefix = "arXiv",
    primaryClass = "hep-ex",
    reportNumber = "CERN-PH-EP-2015-314, LHCB-PAPER-2015-051",
    doi = "10.1007/JHEP02(2016)104",
    journal = "JHEP",
    volume = "02",
    pages = "104",
    year = "2016"
}

@article{Aaij:2020nrf,
    author = "Aaij, Roel and others",
    collaboration = "LHCb",
    title = "{Measurement of $CP$-Averaged Observables in the $B^{0}\rightarrow K^{*0}\mu^{+}\mu^{-}$ Decay}",
    eprint = "2003.04831",
    archivePrefix = "arXiv",
    primaryClass = "hep-ex",
    reportNumber = "LHCb-PAPER-2020-002, CERN-EP-2020-027",
    doi = "10.1103/PhysRevLett.125.011802",
    journal = "Phys. Rev. Lett.",
    volume = "125",
    number = "1",
    pages = "011802",
    year = "2020"
}

@Article{PERF-2007-01,
    author         = "{ATLAS Collaboration}",
    title          = "{The ATLAS Experiment at the CERN Large Hadron Collider}",
    journal        = "JINST",
    volume         = "3",
    year           = "2008",
    pages          = "S08003",
    doi            = "10.1088/1748-0221/3/08/S08003",
    primaryClass   = "hep-ex",
}

@Article{SOFT-2010-01,
    author         = "{ATLAS Collaboration}",
    title          = "{The ATLAS Simulation Infrastructure}",
    journal        = "Eur. Phys. J. C",
    volume         = "70",
    year           = "2010",
    pages          = "823",
    doi            = "10.1140/epjc/s10052-010-1429-9",
    eprint         = "1005.4568",
    archivePrefix  = "arXiv",
    primaryClass   = "physics.ins-det",
}

@Article{EXOT-2012-23,
    author         = "{ATLAS Collaboration}",
    title          = "{Search for high-mass dilepton resonances in \(pp\) collisions at \(\sqrt{s} = 8\,\text{TeV}\) with the ATLAS detector}",
    journal        = "Phys. Rev. D",
    volume         = "90",
    year           = "2014",
    pages          = "052005",
    doi            = "10.1103/PhysRevD.90.052005",
    reportNumber   = "CERN-PH-EP-2014-053",
    eprint         = "1405.4123",
    archivePrefix  = "arXiv",
    primaryClass   = "hep-ex",
}

@Article{PERF-2012-04,
    author         = "{ATLAS Collaboration}",
    title          = "{Performance of \(b\)-jet identification in the ATLAS experiment}",
    journal        = "JINST",
    volume         = "11",
    year           = "2016",
    pages          = "P04008",
    doi            = "10.1088/1748-0221/11/04/P04008",
    reportNumber   = "CERN-PH-EP-2015-216",
    eprint         = "1512.01094",
    archivePrefix  = "arXiv",
    primaryClass   = "hep-ex",
}

@Article{STDM-2012-23,
    author         = "{ATLAS Collaboration}",
    title          = "{Measurement of the \(Z/\gamma^*\) boson transverse momentum distribution in \(pp\) collisions at \(\sqrt{s} = 7\,\text{TeV}\) with the ATLAS detector}",
    journal        = "JHEP",
    volume         = "09",
    year           = "2014",
    pages          = "145",
    doi            = "10.1007/JHEP09(2014)145",
    reportNumber   = "CERN-PH-EP-2014-075",
    eprint         = "1406.3660",
    archivePrefix  = "arXiv",
    primaryClass   = "hep-ex",
}

@Article{PERF-2014-03,
    author         = "{ATLAS Collaboration}",
    title          = "{Performance of pile-up mitigation techniques for jets in \(pp\) collisions at \(\sqrt{s} = 8\,\text{TeV}\) using the ATLAS detector}",
    journal        = "Eur. Phys. J. C",
    volume         = "76",
    year           = "2016",
    pages          = "581",
    doi            = "10.1140/epjc/s10052-016-4395-z",
    reportNumber   = "CERN-PH-EP-2015-206",
    eprint         = "1510.03823",
    archivePrefix  = "arXiv",
    primaryClass   = "hep-ex",
}

@Article{PERF-2014-07,
    author         = "{ATLAS Collaboration}",
    title          = "{Topological cell clustering in the ATLAS calorimeters and its performance in LHC Run~1}",
    journal        = "Eur. Phys. J. C",
    volume         = "77",
    year           = "2017",
    pages          = "490",
    doi            = "10.1140/epjc/s10052-017-5004-5",
    reportNumber   = "CERN-PH-EP-2015-304",
    eprint         = "1603.02934",
    archivePrefix  = "arXiv",
    primaryClass   = "hep-ex",
}

@Article{PERF-2015-09,
    author         = "{ATLAS Collaboration}",
    title          = "{Jet reconstruction and performance using particle flow with the ATLAS Detector}",
    journal        = "Eur. Phys. J. C",
    volume         = "77",
    year           = "2017",
    pages          = "466",
    doi            = "10.1140/epjc/s10052-017-5031-2",
    reportNumber   = "CERN-EP-2017-024",
    eprint         = "1703.10485",
    archivePrefix  = "arXiv",
    primaryClass   = "hep-ex",
}

@Article{PERF-2015-10,
    author         = "{ATLAS Collaboration}",
    title          = "{Muon reconstruction performance of the ATLAS detector in proton--proton collision data at \(\sqrt{s} = 13\,\text{TeV}\)}",
    journal        = "Eur. Phys. J. C",
    volume         = "76",
    year           = "2016",
    pages          = "292",
    doi            = "10.1140/epjc/s10052-016-4120-y",
    reportNumber   = "CERN-EP-2016-033",
    eprint         = "1603.05598",
    archivePrefix  = "arXiv",
    primaryClass   = "hep-ex",
}

@Article{EXOT-2016-05,
    author         = "{ATLAS Collaboration}",
    title          = "{Search for new high-mass phenomena in the dilepton final state using \(36\,\text{fb}^{-1}\) of proton--proton collision data at \(\sqrt{s} = 13\,\text{TeV}\) with the ATLAS detector}",
    journal        = "JHEP",
    volume         = "10",
    year           = "2017",
    pages          = "182",
    doi            = "10.1007/JHEP10(2017)182",
    reportNumber   = "CERN-EP-2017-119",
    eprint         = "1707.02424",
    archivePrefix  = "arXiv",
    primaryClass   = "hep-ex",
}

@Article{EGAM-2018-01,
    author         = "{ATLAS Collaboration}",
    title          = "{Electron and photon performance measurements with the ATLAS detector using the 2015--2017 LHC proton--proton collision data}",
    journal        = "JINST",
    volume         = "14",
    year           = "2019",
    pages          = "P12006",
    doi            = "10.1088/1748-0221/14/12/P12006",
    reportNumber   = "CERN-EP-2019-145",
    eprint         = "1908.00005",
    archivePrefix  = "arXiv",
    primaryClass   = "hep-ex",
}

@Article{EXOT-2018-30,
    author         = "{ATLAS Collaboration}",
    title          = "{Search for a heavy charged boson in events with a charged lepton and missing transverse momentum from \(pp\) collisions at \(\sqrt{s} = 13\,\text{TeV}\) with the ATLAS detector}",
    journal        = "Phys. Rev. D",
    volume         = "100",
    year           = "2019",
    pages          = "052013",
    doi            = "10.1103/PhysRevD.100.052013",
    reportNumber   = "CERN-EP-2019-100",
    eprint         = "1906.05609",
    archivePrefix  = "arXiv",
    primaryClass   = "hep-ex",
}

@Article{FTAG-2018-01,
    author         = "{ATLAS Collaboration}",
    title          = "{ATLAS \(b\)-jet identification performance and efficiency measurement with \(t\bar{t}\) events in \(pp\) collisions at \(\sqrt{s} = 13\,\text{TeV}\)}",
    journal        = "Eur. Phys. J. C",
    volume         = "79",
    year           = "2019",
    pages          = "970",
    doi            = "10.1140/epjc/s10052-019-7450-8",
    reportNumber   = "CERN-EP-2019-132",
    eprint         = "1907.05120",
    archivePrefix  = "arXiv",
    primaryClass   = "hep-ex",
}

@Article{JETM-2018-05,
    author         = "{ATLAS Collaboration}",
    title          = "{Jet energy scale and resolution measured in proton--proton collisions at \(\sqrt{s} = 13\,\text{TeV}\) with the ATLAS detector}",
    year           = "2020",
    reportNumber   = "CERN-EP-2020-083",
    eprint         = "2007.02645",
    archivePrefix  = "arXiv",
    primaryClass   = "hep-ex",
}

@Article{MUON-2018-03,
    author         = "{ATLAS Collaboration}",
    title          = "{Muon reconstruction and identification efficiency in ATLAS using the full Run~2 \(pp\) collision data set at \(\sqrt{s} = 13\,\text{TeV}\)}",
    year           = "2020",
    reportNumber   = "CERN Preprint ID: CERN-EP-2020-199",
    eprint         = "2012.00578",
    archivePrefix  = "arXiv",
    primaryClass   = "hep-ex",
}

@Article{PIX-2018-001,
    author         = "Abbott, B. and others",
    title          = "{Production and integration of the ATLAS Insertable B-Layer}",
    journal        = "JINST",
    volume         = "13",
    year           = "2018",
    pages          = "T05008",
    doi            = "10.1088/1748-0221/13/05/T05008",
    eprint         = "1803.00844",
    archivePrefix  = "arXiv",
    primaryClass   = "physics.ins-det",
}

@Article{TRIG-2018-01,
    author         = "{ATLAS Collaboration}",
    title          = "{Performance of the ATLAS muon triggers in Run~2}",
    journal        = "JINST",
    volume         = "15",
    year           = "2020",
    pages          = "P09015",
    doi            = "10.1088/1748-0221/15/09/p09015",
    reportNumber   = "CERN-EP-2020-031",
    eprint         = "2004.13447",
    archivePrefix  = "arXiv",
    primaryClass   = "hep-ex",
}

@Article{TRIG-2018-05,
    author         = "{ATLAS Collaboration}",
    title          = "{Performance of electron and photon triggers in ATLAS during LHC Run~2}",
    journal        = "Eur. Phys. J. C",
    volume         = "80",
    year           = "2020",
    pages          = "47",
    doi            = "10.1140/epjc/s10052-019-7500-2",
    reportNumber   = "CERN-EP-2019-169",
    eprint         = "1909.00761",
    archivePrefix  = "arXiv",
    primaryClass   = "hep-ex",
}

@Booklet{ATL-SOFT-PUB-2020-001,
    author         = "{ATLAS Collaboration}",
    title          = "{ATLAS Computing Acknowledgements}",
    howpublished   = "{ATL-SOFT-PUB-2020-001}",
    url            = "https://cds.cern.ch/record/2717821",
}

@Report{ATLAS-TDR-2010-19,
    author         = "{ATLAS Collaboration}",
    title          = "{ATLAS Insertable B-Layer Technical Design Report}",
    type           = "ATLAS-TDR-19; CERN-LHCC-2010-013",
    year           = "2010",
    url            = "https://cds.cern.ch/record/1291633",
    related        = "ATLAS-TDR-2010-19-addm",
    relatedstring  = "Addendum:",
}

@Booklet{ATLAS-CONF-2019-021,
    author         = "{ATLAS Collaboration}",
    title          = "{Luminosity determination in \(pp\) collisions at \(\sqrt{s} = 13\,\text{TeV}\) using the ATLAS detector at the LHC}",
    howpublished   = "{ATLAS-CONF-2019-021}",
    url            = "https://cds.cern.ch/record/2677054",
    year           = "2019",
}

@Booklet{ATL-PHYS-PUB-2010-013,
    author         = "{ATLAS Collaboration}",
    title          = "{The simulation principle and performance of the ATLAS fast calorimeter simulation FastCaloSim}",
    howpublished   = "{ATL-PHYS-PUB-2010-013}",
    url            = "https://cds.cern.ch/record/1300517",
    year           = "2010",
}

\end{document}